\documentclass[acmsmall]{acmart}

\usepackage{algorithmic}
\usepackage[shortlabels]{enumitem}
\usepackage{graphicx}
\usepackage{textcomp}
\usepackage{xcolor}
\usepackage{tcolorbox}
\usepackage{balance}
\usepackage{listings}
\usepackage[T1]{fontenc}
\usepackage{beramono}
\usepackage{tikz}
\usepackage{hyperref}
\usepackage{url}
\usepackage{makecell}
\usepackage{lscape}
\usepackage{multirow}
\usepackage[flushleft]{threeparttable}
\usepackage{afterpage}
\usepackage{subcaption}
\usepackage{setspace}
\usepackage[autostyle,german=guillemets]{csquotes}

\SetBlockEnvironment{innerquote}

\renewcommand{\mkbegdispquote}[2]{\textooquote}

\definecolor{blue}{HTML}{000000}
\definecolor{violet}{HTML}{000000}
\definecolor{teal}{HTML}{000000}
\definecolor{red}{HTML}{000000}
\definecolor{major-rev}{HTML}{000000}
\definecolor{minor-rev}{HTML}{000000}

\AtBeginDocument{%
  \providecommand\BibTeX{{%
    \normalfont B\kern-0.5em{\scshape i\kern-0.25em b}\kern-0.8em\TeX}}}

\setcopyright{acmcopyright}
\copyrightyear{2024}
\acmYear{2024}
\acmDOI{XXXXXXX.XXXXXXX}

\begin{document}

\title{Explaining Explanations: An Empirical Study of Explanations in Code Reviews}

\author{Ratnadira Widyasari}
\email{ratnadiraw.2020@phdcs.smu.edu.sg}
\affiliation{%
  \institution{Singapore Management University}
  \city{Singapore}
  \state{Singapore}
  \country{Singapore}
}

\author{Ting Zhang}
\affiliation{%
  \institution{Singapore Management University}
  \city{Singapore}
  \state{Singapore}
  \country{Singapore}
}
\email{tingzhang.2019@phdcs.smu.edu.sg}

\author{Abir Bouraffa}
\affiliation{%
  \institution{University of Hamburg}
  \city{Hamburg}
  \country{Germany}
}
\email{abir.bouraffa@uni-hamburg.de}

\author{Walid Maalej}
\affiliation{%
  \institution{University of Hamburg}
  \city{Hamburg}
  \country{Germany}
}
\email{walid.maalej@uni-hamburg.de}

\author{David Lo}
\email{davidlo@smu.edu.sg}
\affiliation{%
  \institution{Singapore Management University}
  \city{Singapore}
  \state{Singapore}
  \country{Singapore}
}


\begin{abstract}
Code reviews are central for software quality assurance. 
Ideally, reviewers should explain their feedback to enable authors of code changes to understand the feedback and act accordingly. 
Different developers might need different explanations in different contexts. 
Therefore, assisting this process first requires understanding the types of explanations reviewers usually provide. 
The goal of this paper is to study the types of explanations used in code reviews and explore the potential of Large Language Models (LLMs), specifically ChatGPT, in generating these specific types. 
We extracted 793 code review comments from Gerrit and manually labeled them based on whether they contained a suggestion, an explanation, or both. 
Our analysis shows that 42\% of comments only include suggestions without explanations. 
We categorized the explanations into seven distinct types including rule or principle, similar examples, and future implications. 
When measuring their prevalence,  we observed that some explanations are used differently by novice and experienced reviewers. 
Our manual evaluation shows that, when the explanation type is specified, ChatGPT can correctly generate the explanation in 88 out of 90 cases. 
This foundational work highlights the potential for future automation in code reviews, which can assist developers in sharing and obtaining different types of explanations as needed, thereby reducing back-and-forth communication.
\end{abstract}




\maketitle

\section{Introduction}
\label{sec:intro}
Code review is the process of comprehending code changes made by other developers to examine the code content and verify that its quality is sufficient to be merged to the main repository~\cite{morales2015code,tufano2021towards}.
Prior studies have shown the importance of code reviews in software engineering practice~\cite{bavota2015four,morales2015code,mcintosh2014impact}.
Bavota et al.~\cite{bavota2015four}, e.g., found that unreviewed commits have more than twice the chance of introducing bugs and are less readable compared to reviewed commits.
Morales et al.~\cite{morales2015code} found that good code review practices could help improve the design and quality of software systems.

A successful code review requires active participation and collaboration from both \textit{reviewers} and code change \textit{authors}~\cite{sadowski2018modern}.
While the literature has extensively discussed the importance of \textbf{explaining} code changes to facilitate reviewers' work~\cite{pascarella2018information,kononenko2016code}, relatively little attention has been given to elucidating the code reviews themselves and their role in aiding authors to comprehend feedback and minimize communication overhead. 
It is important for code reviewers to explain their decision to help the author understand their review comments~\cite{google_guide, rahman2022example}. 
Rahman et al.~\cite{rahman2022example} recently suggested recommending similar explanations in code reviews to help authors understand feedback.
However, there is still a lack of comprehensive research of how reviewers provide explanations during code reviews and the types of explanations they usually use.

Moreover, recent advance with Large Language Models (LLMs) has shown promising results in assisting developers to reason about code \cite{macneil2023experiences}---potentially reducing the code review effort.
As LLMs are getting widely adopted, there is a growing emphasis on the quality of explanations they provide and on tailoring the explanations to the specific needs of human recipients.
To better understand how humans create explanations, research in the field of eXplainable Artificial Intelligence (XAI) has built on knowledge from philosophy and cognitive psychology, as exemplified by works of Wang et al.~\cite{wang2019designing} and Lim et al.~\cite{lim2019these}.
However, it remains unclear how a targeted usage of LLMs for the code review explanations should look like.
To fill the gap, this work aims to study the types of explanations in code reviews and explore the potential of LLMs in generating specific explanations.

\textbf{Our Work.} To understand how developers usually provide explanations in code reviews, we first crawled Gerrit to collect code reviews from eight popular projects: Android, Bazel,
Chromium, Dart, Flutter, Fuschia, Gerrit, and Go.
Following Bosu et al.~\cite{bosu2015characteristics}, we filtered the reviews to include only those deemed useful. 
We then analyzed the reviewers' comments in useful reviews to extract the explanations and suggestions therein. 
Our analysis revealed that a considerable proportion of the first code review comments (42\%) only include suggestions without  explanations. 
Meanwhile, the code reviews that offer an explanation are typically accompanied by a suggestion (79\% of cases). 
Furthermore, we manually categorized the extracted explanations using the open card sorting technique. 
Based on the pattern of explanation used by the reviewers to clarify and justify their feedback we identified seven categories.
Of these, the most prevalent explanations are {\em stating the cause of an issue}  (Category 6) and {\em providing a benefit of applying the reviewer's suggestion} (Category 7). 
They appear in 40\% resp.~25\% of the code reviews with explanations. 

When investigating the effect of reviewers' experience on the  explanation types provided, we discovered statistically significant differences between novice and experienced reviewers. 
Experienced reviewers tend to use a broader range of explanations, extending beyond Category 6 ({\em stating the cause of an issue}). 
Additionally, we found that experienced reviewers are more likely to explain by discussing potential future implications than novices are. 
To explore LLMs' ability to adapt their explanation to developers' needs, we created a set of ChatGPT prompts, each generating a specific type of explanation given the location of a suspected issue, code, and original code review comment. 
The results show that ChatGPT can successfully transform the explanation to match the intended type in 87 out of 90 cases. 
Additionally, the explanation generated by ChatGPT demonstrates a high level of clarity and informativeness. 

The contribution of our study is fourfold:
\begin{itemize}
    \item An analysis of how often coder reviewers include explanations in their code review comments.
    \item A categorization of reviewer explanations in code review comments.
    \item An analysis of the impact of reviewers' experience on the frequency and type of explanations in code review comments.
    \item A manual evaluation of ChatGPT's ability to generate specific types of review explanations.
\end{itemize}

Since knowledge sharing is key to the code review process and constitutes a significant benefit beyond quality assurance \cite{bacchelli2013expectations,sadowski2018modern}, we think that one added value of our work lies in understanding how developers communicate their feedback and more generally how knowledge sharing occurs around source code \cite{Maalej:BotSE:2023, maalej2014comprehension}.  
Furthermore, with the growing popularity of LLMs in software development, there is a growing need to make LLMs more proficient at explaining code (including suggested changes) to developers. 
Our study can inform future researchers on improving LLMs’ explanation capabilities to better satisfy the needs of the developers seeking explanations.

The remainder of this paper is organized as follows. 
In Section~\ref{sec:background}, we summarize preliminary studies on code review and its  automation. 
Section~\ref{sec:data} introduces our research methodology including the data collection process and research questions. 
We present the analysis results along the research questions in Section~\ref{sec:results}. 
Then, we discuss the findings in Section~\ref{sec:discussion} with potential  implications and threats to the validity.
Finally, we conclude our work and outline future directions in Section~\ref{sec:conclusion}.

\section{Background and Previous Work}
\label{sec:background}

\subsection{Code Review} 

Code reviews, also referred to as peer reviews, serve to ensure the code base's quality. 
Code review is a manual inspection of source code by a developer (i.e., \textit{reviewer}) other than the \textit{author} of the code to find defects and improve the quality of the code~\cite{bacchelli2013expectations}. 
The code review process starts when the code author uploads the code and creates a review request indicating that the code is ready for review. 
In this step, the code author can also invite the appropriate reviewer for the reviewing process. 
The code reviewer will then review the code by providing comments on the code. 

Several code review tools are used by developers to help with the code review process, such as Gerrit \cite{gerrit}, 
Phabricator \cite{phacility}, 
GitHub pull requests \cite{github}, 
and Code Flow \cite{codeflow}. 
These tools have several features, such as showing the difference (\textit{Diff}) between the previous code and the uploaded code, showing commit history, and enabling inline comments so the reviewer can highlight and give feedback on a specific line of code~\cite{turzo2023makes}. 
After receiving feedback from the reviewer, the author can either modify the files and upload a new revi on or respond to the comments. 
After making changes, the author can request a  new review. When the reviewers are satisfied with the revision, they can approve the change or request further modifications.
This review cycle can be repeated multiple times until the reviewers approve the code or until the author abandons the changes.

Several guides on how to write a good code review emerged, among the most popular is   
The Code Review Guidelines by Google ~\cite{google_guide}.
It highlights the reviewer's role to help the author understand why a review comment is made and to explain the review reasoning. 
This study investigates such explanations based on data from Gerrit, a code review tool developed by Google. 
Gerrit is a central tool at Google used for the development of products built with Git, including Android and Chromium.

\vspace{0.2cm}
\noindent\textbf{Useful Code Reviews.} 
Previous studies investigated the factors that make a code review useful~\cite{hasan2021using, bosu2015characteristics}. 
Bosu at al.~\cite{bosu2015characteristics} conducted an empirical study at Microsoft,  including developers interviews, manually labeling code review data, training an automated model to classify the usefulness of code review comments, and analyzing predictions among 1.5 million code review comments. 
The study identified a set of factors affecting the usefulness of code reviews, such as thread status and the number of comments.
A subsequent study by Rahman et al.~\cite{rahman2017predicting} developed a tool called RevHelper using both textual features and developer experience to predict the usefulness of code review comments.
While Bosu et al.'s approach~\cite{bosu2015characteristics} focuses on classifying the usefulness of a comment post-review completion, RevHelper by Rahman et al. ~\cite{rahman2017predicting} predicts the usefulness of a code review comment
as soon as it is made using only pre-completion attributes of a review.
RevHelper achieves 66\% accuracy ~\cite{rahman2017predicting} and outperforms three variants of Bosu et al.'s model using only a subset of the features available prior to code review completion.
Another study by Hasan et al.~\cite{hasan2021using} replicated the interview and manual labeling from Bosu et al.'s study in a different commercial organization.
Their code review classification model combines features from both previous works by Bosu et al.~\cite{bosu2015characteristics} and Rahman et al.~\cite{rahman2017predicting} except the features \textit{Thread Status} and \textit{External Library Experience}.
Their model achieves 87.32\% accuracy on the chosen data set.
The results of their interviews showed that developers agree on several usefulness criteria, such as defect identification and solution approach.

\begin{figure*}
\centering
\includegraphics[width=\linewidth]{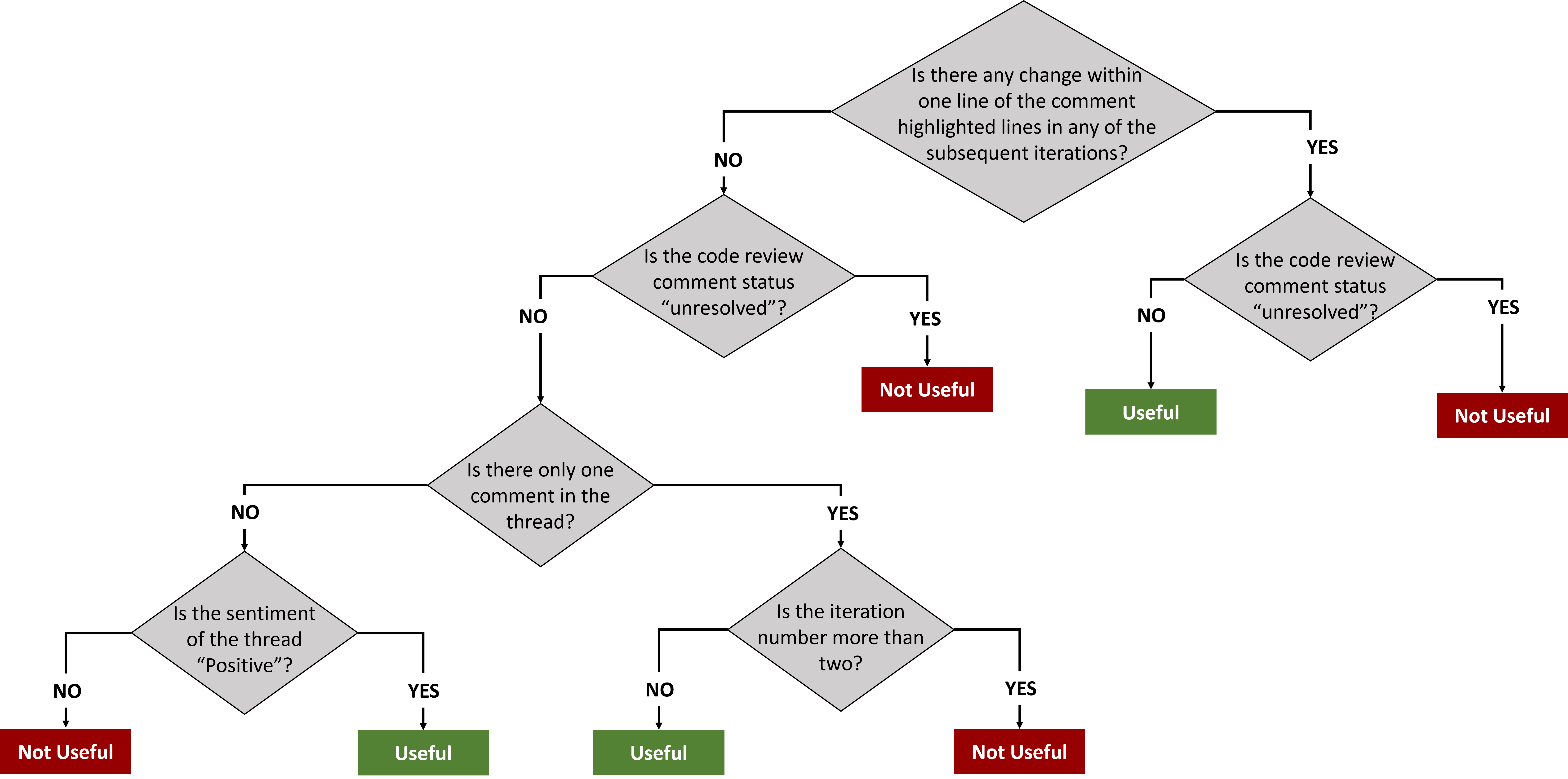}
\caption{Decision tree for useful code review based on Bosu et al.~\cite{bosu2015characteristics} study.}
\label{fig:decision_tree}
\end{figure*}

To identify the useful code reviews in our analysis, we use the decision tree of Bosu et al.~\cite{bosu2015characteristics} as shown in Figure \ref{fig:decision_tree}.
We chose this approach because it is grounded in empirical findings and has been used to predict the usefulness of 1.5 million code review comments. Furthermore, the approach has been used in other studies to classify comment usefulness after the code review completion rather than predicting usefulness during the code review process. 
This decision tree has components such as checking the code review comment status (resolved or unresolved comments), sentiment (obtained through sentiment analysis), and code changes after the review.

\subsection{Explanation in Code Review}
Code review is a collaborative process in which the communication between authors and reviewers plays a central role in achieving a common understanding of the change.
During code review discussions, authors are invited to provide an explanation of their patch if needed.
A study by Liang et al.~\cite{liang2019explain} investigated how developers explain bug-fixing patches in pull requests and found that patch explanations contained different expressive forms.
For instance, developers described either a missing or a wrong process to express the cause of a bug.
However, to ensure that the submitted patch meets the project's quality criteria, it is desirable that reviewers make their feedback understandable and actionable to the authors.
Previous works, e.g.~by Gon\c{c}alves et al. \cite{wurzel2023competencies, wurzel2022interpersonal}, have highlighted the importance of constructive feedback in code reviews.

Constructive feedback relies on the provision of explanations related to various aspects, such as reasons for decisions, what is perceived as wrong, as well as what to do and how to do it.
Rahman et al.~\cite{rahman2022example} proposed an automated approach for example-driven code review explanations recommending similar reviews to help reduce communication overhead. 
The authors labeled 3,722 code reviews for clarity and used the data to train a classifier. Following the prediction, information retrieval techniques were employed to retrieve the top five most similar reviews with higher clarity. 
Unlike previous studies that have mainly focused on enhancing the clarity and usefulness of code reviews in a broad sense, our study specifically investigates the suggestions and explanations by reviewers (i.e. \underline{what} the reviewer thinks should be changed vs.~\underline{why}). 
We aim at exploring the various types of explanations provided by experienced and novice reviewers, aiming to better understand how developers convey their reasoning within the review process.

\begin{figure}
\centering
\includegraphics[scale=0.42]{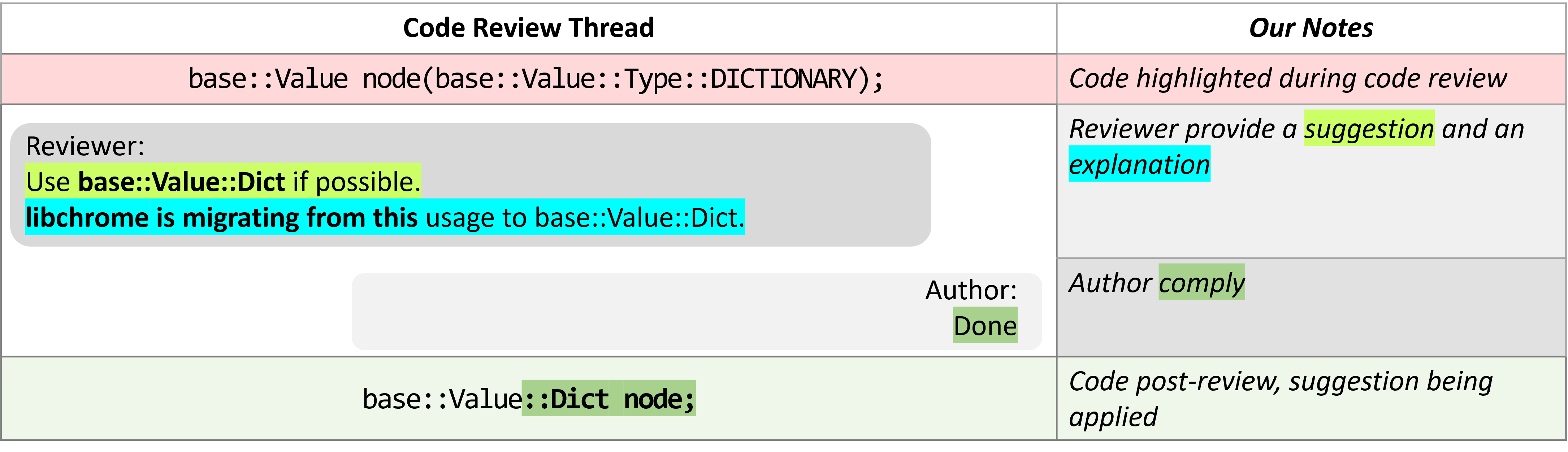}
\caption{Code review example (Chromium project ID-3935924).}
\label{fig:motivation_additional}
\end{figure}

\begin{figure}
\centering
\includegraphics[scale=0.42]{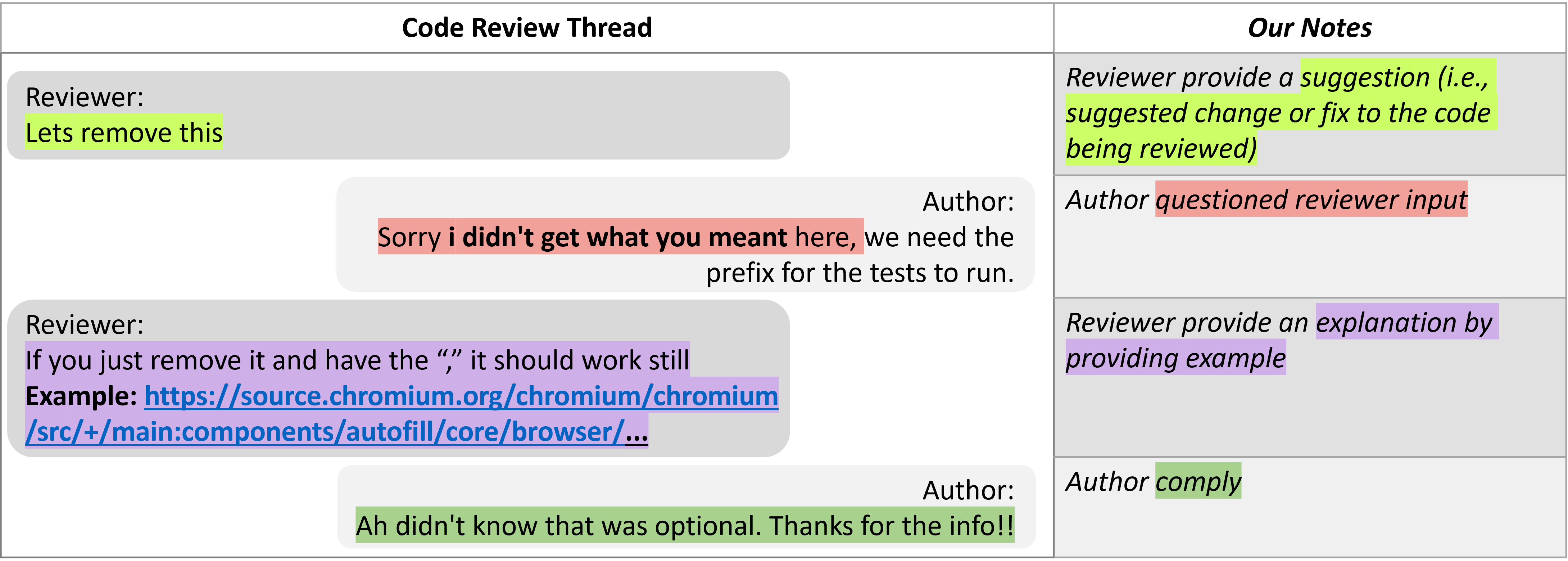}
\caption{Example of a reviewer explanation (Chromium project ID-4585015).}
\label{fig:motivation_1}
\end{figure}

\vspace{0.2cm}
\noindent{\textbf{Motivating Examples.}}\label{sec:motivation}

In our study, we differentiate between suggestions and explanations that reviewers provide in their comments. 
For instance, consider the review comment in Figure ~\ref{fig:motivation_additional}: ``Use ‘base::Value::Dict’ if possible. libchrome is migrating from this usage to ‘base::Value::Dict’.'' 
Here, the directive ``Use ‘base::Value::Dict’ if possible'' represents the suggestion, while ``libchrome is migrating from this usage to ‘base::Value::Dict’,'' serves as the explanation/reason. 
This distinction is crucial as it addresses not just what should be changed but why it should be changed.

A second example presented in Figure~\ref{fig:motivation_1} underscores the role of such explanations. 
The reviewer proposed a suggestion (i.e., a modification of the reviewed code) without providing any rationale. 
They simply state ``Let's remove this.'' 
In reply, the author expressed confusion and sought clarification on the necessity of the suggested change. 
Subsequently, the reviewer elaborated on the reasons by providing a similar example. 
The author then understood why the suggested changes were necessary and subsequently implemented the feedback. 
This highlights the importance of providing explanations in code review comments. Especially in complex scenarios, simply asking for a change may fall short of fostering understanding and successful collaboration.

\begin{figure}
\centering
\includegraphics[scale=0.42]{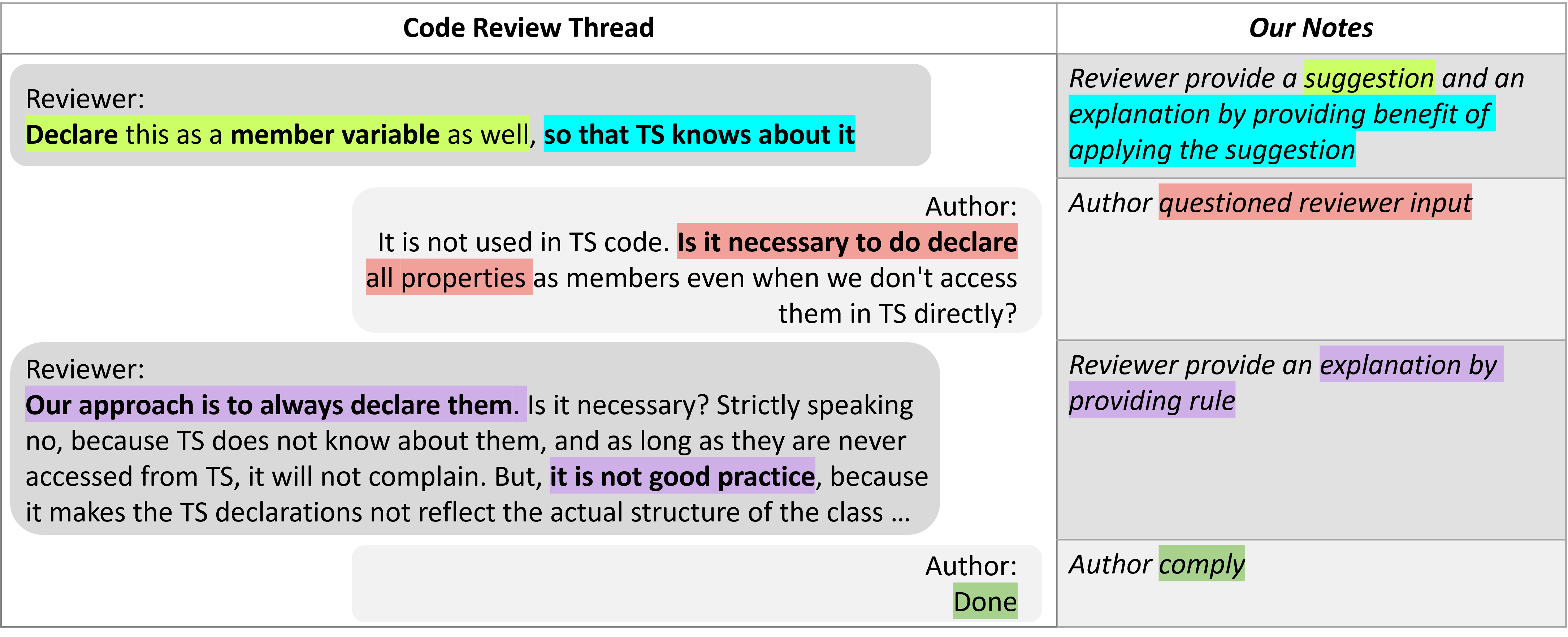}
\caption{Example of different types of explanation by a reviewer in one thread (Chromium project ID-4614863).}
\label{fig:motivation_2}
\end{figure}

Furthermore, there are several ways for the reviewer to explain why a change is needed. 
Figure~\ref{fig:motivation_2} highlights the impact of different explanations on the author's response. The reviewer initially proposed a change that she thinks is needed (i.e., suggestion). 
Then she gave an explanation by providing an implication or benefit if the author followed the suggestion, stating ``Declare this as a member variable as well, so that TS knows about it.'' Seeking clarification, the author questioned the necessity of this change. 
In response, the reviewer cited the project rule of always declaring member variables and emphasized that not declaring the variable goes against good programming practice. 
Once the reviewer highlighted this rule in their explanation, the author eventually accepted and implemented the feedback. 
This example demonstrates that different types of explanations can lead to different responses by the code author. 
Introducing the needed type of explanation early in the process may save clarification iterations and enhance the efficiency of the code review. 
As Figure~\ref{fig:motivation_2} shows, the first explanation (providing the benefit of applying the suggestion) has resulted in a different response by the author compared to the second explanation (providing a rule or principle). Generating or suggesting different types of explanations might thus benefit both code reviewers and authors, facilitating a more effective and productive code review process. 

\subsection{Code Review Recommendation and Automation}
To facilitate code reviews, several studies attempted to automate aspects of the code review process. 
One notable example is the study by Thongtanunam et al.~\cite{thongtanunam2015should}, which proposed RevFinder that can recommend the most suitable code-reviewer based on the reviewed file location. 
Similarly, Zanjani et al.~\cite{zanjani2015automatically} proposed cHRev to recommend code reviewers based on their historical contributions. 

Other studies focused on the transformation of code reviews, i.e.~automatically transforming a method from a before version (i.e., the version before implementing code review feedback) to an after version (i.e., the version after code review feedback is implemented). 
Tufano et al.~\cite{tufano2019learning} proposed a Neural Machine Translation
(NMT) approach to solve this problem. Tufano et al.~\cite{tufano2021towards} also explored the pre-trained models for code review automation. 
In a subsequent study~\cite{tufano2022using}, the authors introduced deep learning models that automate code change recommendations during code reviews. 
Thongtanunam et al.~\cite{thongtanunam2022autotransform} improve the previous study by proposing AutoTransform which utilizes a Byte-Pair Encoding (BPE) and Transformer-based NMT architecture to handle new tokens and long sequences.

Other researchers have delved into assisting code review comments. 
For instance, Gupta and Sundaresan~\cite{gupta2018intelligent} introduced DeepCodeReviewer, which employs historical peer reviews and deep learning to suggest code review comments linked to common issues.  
Hong et al.~\cite{hong2022commentfinder} proposed CommentFinder an information retrieval (IR) based approach to suggest code review comments by retrieving similar code that has been previously reviewed.
Li et al.~\cite{li2022automating} used pre-training techniques in three code review activities (code change quality estimation, code review comment generation, and code refinement). 
Furthermore, Li et al.~\cite{li2022auger} proposed AUGER (AUtomatically GEnerating Review comments) that uses pre-trained models to generate code review comments.

In this study, we also explore the potential use of Large Language Models (LLMs) to provide assistance during the code review process. 
LLMs and particularly ChatGPT have been widely applied in various studies within the Software Engineering (SE) domain~\cite{liu2023refining,liu2024your,guo2024exploring,sun2024gptscan, zhou2024large}, such as code generation~\cite{liu2024your, liu2023refining},  vulnerability detection~\cite{sun2024gptscan, zhou2024large}, and fault localization~\cite{jiang2024evaluating, widyasari2024demystifying}. 
In relation to code review, a previous study by Guo et al.~\cite{guo2024exploring} used ChatGPT for code refinement based on code review comments.

We chose ChatGPT, specifically GPT3.5, due to its popularity, top performance, and widespread accessibility compared to other models (e.g., LLaMA ~\cite{Meta_LLaMA3}, Gemini~\cite{Google_Gemini}, and Mixtral~\cite{Mistral_Mixtral}) especially for its applications in the SE domain~\cite{liu2023refining, liu2024your,guo2024exploring,jiang2024evaluating,zhou2024large}.
Based on our work, developers can customize the explanations provided by reviewers to their preferred types. 
For instance, if the authors of the code change (i.e., those reading the code review comments) prefer explanations that provide similar examples, the LLM can transform the review comments to fit this category. Tailoring explanations to developer’s preferences and needs can lead to more effective and efficient code reviews.

\section{Methodology}
\label{sec:data}
\subsection{Research Questions}
This work aims at answering the following research questions (RQs):

\textbf{RQ1: How often do reviewers include explanations in code reviews?}
Our objective is to determine how often reviewers provide reasons or justifications for their code review comments. While it has been repeatedly highlighted that code reviewers need to provide reasoning in their comments to help authors understand the feedback and implement necessary changes effectively, the extent to which reviewers actually include such reasoning in practice is unknown. An explanation may include the status of the current code\footnote{\url{https://gerrit-review.googlesource.com/c/gerrit/+/345833/7..10/java/com/google/gerrit/server/index/change/ChangeField.java\#b1407}}, a condition or scenario in which it fails\footnote{\url{https://gerrit-review.googlesource.com/c/gerrit/+/345833/5..10/java/com/google/gerrit/server/index/change/ChangeField.java\#b1403}} among other possible facts used by reviewers to explain their feedback. 
The Cambridge English dictionary defines an explanation as \textit{``the details or reasons that someone gives to make something clear or easy to understand''}\footnote{https://dictionary.cambridge.org/dictionary/english/explanation}.
To the best of our knowledge, there are no previous studies that have formally described `explanation' in the code review context. Considering these justifications and reasons as forms of explanation and combining this with a general description of `explanation', we describe an explanation in the code review comment as the rationale behind why the code being reviewed needs to be changed. Furthermore, we also check whether the reviewer includes a suggestion in their comment. We describe a suggestion in a code review comment as the actual code change or recommendation proposed by the reviewer to address an issue or improve the existing code. An example of a code review comment having both an explanation and a suggestion is shown in Table~\ref{tab:example_label_explanation} C6. 
In this example, the explanation highlights that the author needs to change the method name \textit{power\_button} as this variable name does not exist in the code.
An example of a reviewer suggestion is also shown in Table~\ref{tab:example_label_explanation} C6, which is to change the variable name to ``PowerButton".

\textbf{RQ2: What type of explanations are found in code review comments?}
For the second research question, our goal is to identify the types of explanations that reviewers use in their code review comments. Understanding the different types of explanations is crucial because it provides insights into how reviewers convey their feedback and reasoning to code authors. By categorizing these explanations, we can better understand the communication patterns in code reviews and identify which types are most commonly used in which context.

\textbf{RQ3: How does reviewers’ experience affect the frequency and type of explanations?}
Since code review is a human-centered activity in which the reviewers' experience may play a significant role~\cite{rahman2017predicting, kononenko2015investigating, kononenko2016code, uchoa2021predicting}, we further analyzed the results of the first and second RQs taking into account the reviewers' experience. 
Experienced reviewers may provide different, more detailed, or frequent explanations due to a deeper understanding of the codebase and certain communication skills. 
In contrast, less experienced reviewers might rely on different types of explanations. 
In this RQ, we aim to deepen our analysis of reviewer explanations by investigating potential differences in explanations based on experience. This is important because identifying such differences can guide strategies for better communication and collaboration among developers. 
It can also help less experienced reviewers improve their code review comments.

\textbf{RQ4: To what extent can ChatGPT generate a specific type of explanation?}
ChatGPT's widespread adoption is due to its unprecedented performance on a wide range of natural language generation tasks. 
While the full range of tasks that ChatGPT can assist developers with may not have been fully explored yet, the developer community has been eager to test the full extent of its capabilities in software development. 
In this study, we explored a potential use case for leveraging various types of explanations identified in RQ2 by utilizing ChatGPT. Specifically, we used ChatGPT to transform one type of code review explanation into another. To generate the explanation, we provided ChatGPT with specific prompts, each accompanied by several pieces of information. This generation of different types of explanations in code review comments using ChatGPT aims to support developers in their code review process.

\subsection{Method}
\begin{figure}
\centering
\includegraphics[scale=0.42]{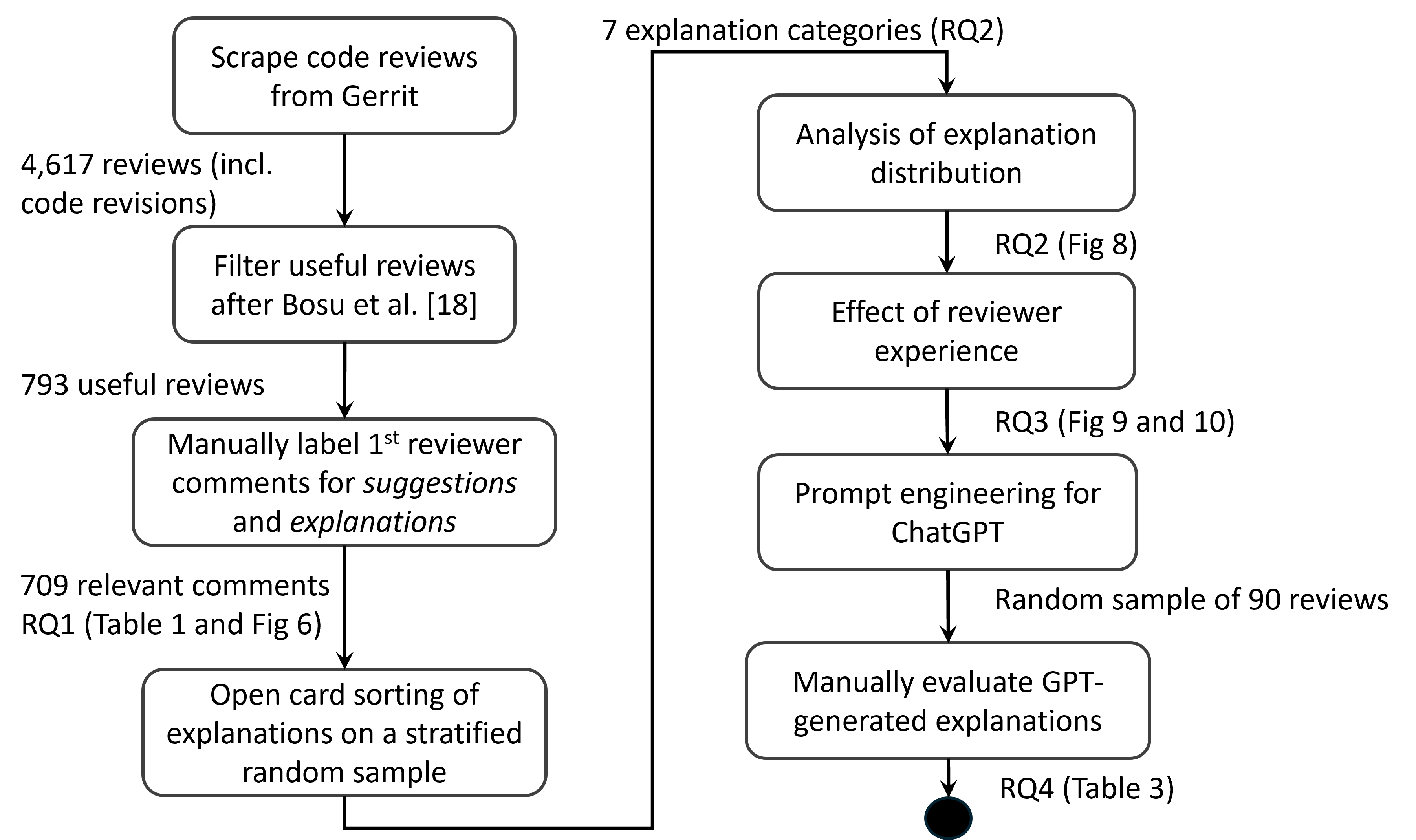}
\caption{Overview of our research methodology.}
\label{fig:methodology}
\end{figure}

In this section, we describe the methodology of our study. An overview of the methodology is presented in Figure~\ref{fig:methodology}. First, we collected the dataset for our study by scraping Google projects on Gerrit and then filtering useful code review comments. 
Our first research question (RQ1) investigates the frequency with which reviewers justify their decisions during code reviews. 
To address this, we manually categorized the reviewers' comments into suggestions and explanations. For RQ2, we aimed to identify the types of explanations present in code review comments. We employed open card sorting on a randomly stratified sample and subsequently analyzed the results. 
In RQ3, we examined the effect of developers' experience on the frequency and types of explanations used. 
Lastly, for RQ4, we investigated how ChatGPT can be utilized to transform specific explanations into different types. This involved prompt engineering with manual evaluation to derive a best-performing prompt, followed by an extended evaluation of explanations generated with this prompt.

\subsubsection{Data Collection}
\label{sec:dataset}

We collected a total of 4,617 reviews with 9,169 revisions from Google open source projects available on Gerrit \cite{gerrit_cr} during the period from October 5 to 9, 2022. The project listings were obtained from the Google open-source project page.\footnote{\url{https://opensource.google/projects}} We opted for Google open source projects because Google provides a guide that emphasizes the need for reviewers to explain their review decisions thoroughly~\cite{google_guide}. The Google open-source projects we focused on were those accessible on Gerrit, specifically projects that make the \url{https://[project_name]-review.googlesource.com} domain available to the public. These projects included Android~\cite{android_cr}, Bazel~\cite{bazel_cr}, Chromium~\cite{chromium_cr}, Dart~\cite{dart_cr}, Flutter~\cite{flutter_cr}, Fuschia~\cite{fuschia_cr}, Gerrit~\cite{gerrit_cr}, and Go~\cite{go_cr}. 
We revised the implementation published by Paixao et al.~\cite{paixao2018crop} to retrieve up-to-date Gerrit code reviews.

Inspired by prior studies, we focussed on analyzing code review comments considered useful to ensure the data that we analysed has a high quality and is relevant for the manual analysis. To determine code review comments usefulness for developers, Bosu et al.~\cite{bosu2015characteristics} conducted an interview study with developers at Microsoft. Their main finding was that developers classified comments as useful if they highlighted issues in the code,  presented alternative scenarios, or suggested different implementations. Nitpicking was also considered somewhat useful, e.g. for long-term project maintenance. On the other hand, interviewees found comments to be not useful if they contained clarification questions, praise or future work not related to the current change under review. The insights from previous study into what constitutes a useful versus a not useful explanation further strengthen our decision to focus on useful code review comments. These are more likely to contain explanations, while not useful comments, which typically ask for clarifications or offer praise, provide little additional information to the code author. For example, a non-useful code review comment such as "lgtm, thanks!"—extracted from data we have scraped from Gerrit—lacks both an explanation and a suggestion for improving the code.

After the interview, Bosu et al.~\cite{bosu2015characteristics} subsequently manually labelled the usefulness of 844 code review comments. From this data, they identified key attributes of useful code review comments, including the status of the comment, whether it triggered a code change in subsequent iterations, and the sentiment expressed in the comment. They then computed these attributes and developed a decision tree model based on them. Their best model showed that the most important features were the change trigger and the status of the comment, as shown in Figure~\ref{fig:decision_tree}. Additionally, other important attributes of the model included the number of comments, iteration number, and the sentiment of the code review.
Since the decision tree was originally built for CodeFlow, an internal code review tool used by Microsoft, we adapted it to be suitable for Gerrit. 
For example, the original decision tree uses code review comment status.
We mapped the ``closed" comment status in CodeFlow to the ``resolved" status in Gerrit.
Moreover, in the original study \cite{bosu2015characteristics}, the authors made use of a sentiment analysis tool presented by Quirk et al.~\cite{quirk2012msr}.
Since this tool is not publicly available, we used the Transformer-based model published by Zhang et al.~\cite{zhang2020sentiment}, outperforming baseline approaches of sentiment analysis in code reviews. They compared and evaluated their method against five baseline methods across six software engineering datasets. The experimental results demonstrated that their method outperformed others by 6.5-35.6\% in terms of macro/micro-averaged F1 scores. 
The replication package~\cite{soarsmu} from Zhang et al.'s work~\cite{zhang2020sentiment} was used for sentiment checking.
The updated decision tree is shown in Figure~\ref{fig:decision_tree}. It is important to note that the methodology for obtaining useful code review comments was not the primary focus of our research. The decision tree used for filtering, although aiding in the extraction of useful comments, does not fundamentally alter the essence of our analysis. The value of our study lies in understanding the content and nature of explanations provided during code reviews.

In addition to filtering by usefulness, we only retain inline code review comments as they offer a clear mapping to the corresponding code sections. Unlike comments on entire code changes or comments at the file level, inline comments offer more targeted insights. 
This selection allowed us to focus on comments that address particular code segments, leading to a more detailed and contextually relevant analysis. 
We identified a total of 1,325 instances of inline comment threads deemed useful. 
The Android project had the highest number of useful inline comments, with 789 instances, while the Dart project had the least, with 50 useful code review comments. 
We note the exclusion of the projects Bazel, Flutter, and Fuschia from the analysis as they did not have any useful comments remaining after the filtering process. Due to the substantially higher number of comments in the Android project compared to all other projects, and to keep the manual analysis reasonably sized, we randomly selected a statistically significant sample for this project. Following previous studies~\cite{gunawardena2023concerns}, we considered a 95\% confidence level with a 5\% margin of error, resulting in a sample size of 259 comments out of the 789 total comments in the Android project. For the other projects, we included all the inline comments resulting from the usefulness filter (Figure~\ref{fig:decision_tree}). Moving forward, we will refer to these inline comments as code review comments to maintain consistency. In total, we have 793 code review comments for further analysis. The final distribution of these code review comment threads across projects is presented in Table~\ref{tab:dist_useful}.

\begin{table}[] 
    \centering
    \caption{Dataset and RQ1 results.}
    \label{tab:dist_useful}
    \vspace{-0.2cm}
\begin{tabular}{|l|r|r|r|r|r|}
\hline
\multirow{2}{*}{\textbf{Project}} & \multirow{2}{*}{\textbf{\#CodeReviewComment}} & \multicolumn{4}{c|}{\textbf{RQ1 Results}}                                                                                 \\ \cline{3-6} 
                           &               & \multicolumn{1}{c|}{\textbf{\#NoExpSugg*}}             & \multicolumn{1}{c|}{\textbf{\#Exp.}\textsuperscript{\textdagger}} & \multicolumn{1}{c|}{\textbf{\#Sugg.}\textsuperscript{\textdaggerdbl}} & \textbf{\#Exp.+Sugg.} \\ \hline

Android          & \textcolor{blue}{259}     
& 33 (13\%) & 21 (8\%)                                        &    115  (44\%)               &    90     (35\%)                            
\\ \hline
Chromium         & 318  & 24 (8\%) & 43  (14\%)                                        & 128  (40\%)               & 121       (38\%)                                                  \\ \hline
Dart             & 50 & 5 (10\%) & 3 (6\%)                                          & 18 (36\%)                 & 24   (48\%)                           \\ \hline
Gerrit           & 53 & 8 (15\%) &  7   (13\%)                                        &   24      (45\%)            &   14      (27\%)                           \\ \hline
Go               & 113 & 12 (11\%) &  7  (6\%)                                         &    46   (41\%)              &  48     (42\%)                       \\ \hline
\textbf{Total} & 793 & 84 (10\%) & 81 (10\%) & 331 (42\%) & 297 (38\%)\\ \hline
\end{tabular}
\\
\raggedright \hspace{0.2cm}
\footnotesize*\textcolor{major-rev}{Refer to the code review comment that does not include neither explanation nor suggestion.} \\ \hspace{0.24cm}\textsuperscript{\textdagger}\textcolor{major-rev}{"Exp.'' referred to the explanation in code review comment.}
\hspace{0.4cm}\textsuperscript{\textdaggerdbl}\textcolor{major-rev}{"Sugg.'' referred to suggestion made by reviewer.}
\vspace{-0.2cm}
\end{table}

\subsubsection{RQ1}
We conducted a manual analysis of the 793 useful code reviews (see Section~\ref{sec:dataset}).
To label a reviewer's comment we identified whether it contains a suggestion, explanation, or both. 
While we read the code line in question as well as the entire discussion thread, we only labelled the first inline code review comment created by the reviewer (Table~\ref{tab:dist_useful}). This initial inline comment is crucial as it typically pertains directly to the reviewed line of code, whereas general comments may address the entire code change~\cite{ebert2017confusion, huq2022review4repair}. Subsequent comments may be from the authors themselves or other developers. They often evolve into more general discussions about the code, which may not be as closely related to the commented code line. Using the first comment ensures that each item in the dataset is comparable (1 representative comment per thread), enhancing consistency. This approach also makes the labeling task more manageable.
Two authors of this paper, each with nine years of programming experience labeled the specific comments. In case of conflicting labels, discussions were held to reach a consensus.

\subsubsection{RQ2}\label{sec:rq2_method} 
For this analysis, we randomly selected a representative number of samples from the code review comments that contained some form of explanation from RQ1 results.
Following the method employed by a previous study by Gunawardena et al.~\cite{gunawardena2023concerns}, we consider a confidence level of 95\% with a 5\% margin of error for the data sampling. 
We used a stratified sampling approach to ensure representation from each project. Our analysis included a total of 298 code review comments containing explanations. The breakdown of the sample, reflective of the population sizes, is as follows: Android (87 comments from a population of 111), Chromium (116 comments from a population of 164), Dart (26 comments from a population of 27), Gerrit (20 comments from a population of 21), and Go (49 comments from a population of 55).

We applied open card sorting~\cite{spencer2009card} as an approach to categorize the explanations present in the first code review comments within the threads.
Open card sorting has been used in many previous studies~\cite{xu2020reinventing, bacchelli2013expectations} for bottom-up analysis and categorization as well as taxonomy creation. 
As the name suggests, open card sorting does not assume pre-defined categories.
Instead, category names are derived a posteriori after all cards have been sorted.

For the card sorting task, we created cards containing information from the reviewer comments, comments in the thread, highlighted code, and code diff from the current version with an updated version (the cards we used can be found in our replication package~\cite{replication_package}). 
Three authors of this paper with nine years of programming experience discussed each card and categorized them into appropriate groups. If there were differing opinions on the categorization, they were discussed and resolved until a unanimous decision was reached. 
At this stage, the cards that were sorted together did not have a group name. 
After all the cards had been sorted, the authors assigned appropriate names to the groups that reflected the patterns of the explanations.
The described open card sorting process is in line with previous studies~\cite{xu2020reinventing, bacchelli2013expectations}.

\subsubsection{RQ3}\label{sec:rq3_additional_method}

Initially, we examined how the experience of developers affects the frequency of explanations given in code review comments. Subsequently, we explored whether the experience of reviewers influences the types of explanations they are more likely to provide. 
We distinguished the reviewers in our dataset into experienced and novices. 
A previous study by Kononenko et al.~\cite{kononenko2016code} computed the experience score by summing the number of submitted and reviewed patches for each reviewer. 
In their study, they used a threshold score of 15, which resulted in 27\% experienced developers. 
Given that the distribution of experience scores in our dataset differs from that of the previous study, directly applying the same threshold may not accurately reflect experience levels. 
To address this, we sorted the reviewers by their scores and labeled the top 27\% as experienced reviewers. 
Furthermore, we opted to use only the number of reviewed patches since our focus is on the explanations provided during code reviews rather than the overall experience of the developer. 
In the end, we registered 67 experienced reviewers and 185 novice reviewers.

\subsubsection{RQ4} \label{sec:rq3_method}
Answering this RQ included two phases: first, a prompt engineering to identify a best-performing prompt, then a manual evaluation with that prompt on a larger sample of comments. 

\textbf{Prompt Engineering and Initial Evaluation. }
In total, this first evaluation comprised 120 code review comments generated from the four prompt variants (detailed in the next paragraph).  Specifically, for each prompt, we generated 30 code review comments covering six different categories (excluding Category 5), with five samples for each category. We excluded \textit{Category 5} (expressing personal preference or opinion of the current code) as ChatGPT was not designed to provide subjective viewpoints as an AI language model. It has minimal impact on the overall results, as Category 5 represents only 9\% of the total dataset.

We carefully designed each prompt to align with the identified explanation type and its corresponding pattern. 
Different information can be included in the prompt such as highlighted line, code, commit message, original code review, and specific instructions for conducting a code review based on a particular type of explanation. 
Following previous studies~\cite{zhou2024large,jiang2024evaluating}, we combined different information to create various prompts. We then manually evaluated the prompts' results to identify the best-performing prompt. 
The base input of the prompt includes the code review comment, code snippet, and the target category. 
The additional inputs are (1) the highlighted code line and (2) the commit message. 
This resulted in four different prompt variations: "Base" (only the base inputs), "Base+Line" (includes both the base and line), "Base+Commit" (includes both the base and commit message), and "All" (includes the base and all additional inputs). 
We did not include the full descriptions of the explanation categories in the prompts, assuming that ChatGPT would understand and generate responses based on the category names alone. To check this assumption, we asked ChatGPT to describe the category of explanation and provide an example for each category using this prompt: ``\textit{As a code reviewer, could you explain the category of explanation in the code review comment called [Category Name]? Please provide an example of a comment in this category.}'' 
This analysis can be found in our replication package along with all other results. 
We observed that ChatGPT can reasonably infer the meaning of each category from its name, which suggests the full category description would not be required as input.

The evaluation of the generated code review comments used two key \textbf{metrics}: the correctness of the explanation type and the correctness of the semantic meaning.
To determine the correctness of the explanation type, evaluators verified whether the generated code review comments belonged to the intended type of explanation. 
For the correctness of the explanation, evaluators compared it with the reviewer explanation that was collected in the previous RQs, ensuring it correctly identified the issue originally pointed out by the code reviewers. 
Additionally, the accuracy of statements provided in ChatGPT-generated explanations needed to be checked, for example, by cross-referencing any rules mentioned with online sources for verification. 
This evaluation process involved two authors of this paper who were not involved in the code review generation. These two authors have nine years of programming experience. 
To mitigate biases, the order of prompt results was randomized and prompt types were removed during the evaluation.

\textbf{Extended Evaluation of Best-Performing Prompt. }
In this extended evaluation phase, the evaluators reviewed a total of 90 code review comments, all generated using the single best-performing prompt. 
By increasing the number of samples to be evaluated (compared to 30 in the initial evaluation), we aimed to provide a more thorough assessment.
The 90 review comments consisted of 15 different samples for each category (excluding the ``personal preference regarding the reviewed code'' category).

Based on the results of the previous step, we  identified the best prompt based on both correctness metrics. 
We used this best-performing prompt to generate additional explanations, which we then manually evaluated in a second round.
In addition to the two correctness \textbf{metrics} used in the first evaluation (explanation type and explanation correctness), the evaluators also assessed the clarity and informativeness of the generated explanations for more extensive evaluation. 
To evaluate clarity and informativeness, we used a 7-level semantic scale, ranging from 1 (Very Unclear) to 7 (Very Clear) for clarity and from 1 (Very Uninformative) to 7 (Very Informative) for informativeness. 
These metrics were chosen based on a previous study~\cite{huang2023ischatgpt} that evaluates the explanation made by ChatGPT for sentiment analysis. 
By incorporating clarity and informativeness metrics in the second evaluation, we aimed to gain a more comprehensive understanding of the quality and effectiveness of the generated explanations.
This step was carried out by five evaluators with an average of 6.4 years programming experience (ranging from four to eight years).  
For both clarity and informativeness, we report the average scores from all five evaluators. 
Since both correctness metrics (the explanation itself and the explanation type) are represented as boolean values, we aggregate the results using a consensus strategy. The generated code review comment is deemed correct if 80\% (four or more) of the five evaluators agree so. We chose the 80\% threshold following previous studies that used consensus approaches ~\cite{rodriguez2013searching,gattrell2024accord}.

\section{Results}
\label{sec:results}
\begin{figure}
\centering
\includegraphics[scale=0.5]{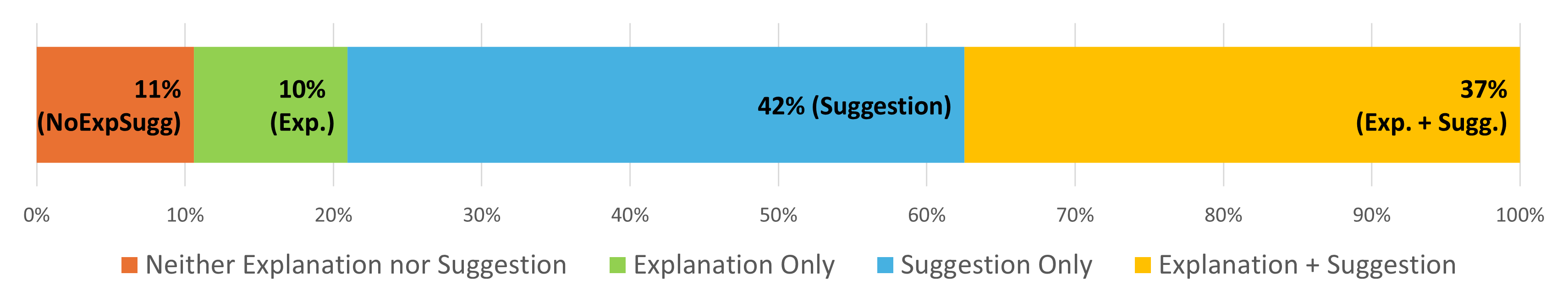}
\vspace{-0.3cm}
\caption{Distribution of explanation, suggestion, and explanation+suggestion. }
\label{fig:venn_rq1}
\vspace{-0.2cm}
\end{figure}

\subsection{RQ1: How Often Do Reviewers Include Explanations in Code Reviews?}

We manually analyzed a total of 793 threads of code review comments introduced in Section~\ref{sec:dataset}.
We assigned a label to the first code review comment in each thread on whether it only had an explanation, a suggestion, or both. An explanation in a code review comment refers to the rationale behind why the code being reviewed needs to be changed, while a suggestion in a code review comment refers to the actual code change or recommendation proposed by the reviewer to address an issue or improve the existing code. 
In our analysis, we also found code review comments where neither explanation nor suggestion was present. 
For example, comments that only mentioned ``same'' and ``ditto'', without highlighting what they were referring to are part of this group. 
Other examples are comments that only asked for confirmation or gave praise such as “When would we reach this now?” or “Nice spot!” without a suggestion or explanation to improve the code.

In the end, we identified 709 code review comments that included some form of explanation or suggestion. Table~\ref{tab:example_label_explanation} shows 6 examples of the labeled code review comments. 
In the first comment C1, we consider it contains a suggestion without explanation as it only contains instructions on how to improve the code, namely by changing the highlighted part of the code from ``EXPECT'' to ``ASSERT''. 
Similarly, in the second example C2, the reviewer asked the author to change ``DupCertBuffer().get()'' to ``GetCertBuffer()''.
In both examples, the reviewer did not give any reason or justification for the proposed change. 
An example of a code review comment only containing an explanation is shown in the third and fourth examples (C3 and C4) in Table~\ref{tab:example_label_explanation}. In C3, the reviewer provides the rationale that the code does not include what's being recorded which points out an issue with the current code without suggesting a specific suggestion. Meanwhile, in C4, the reviewer provides reasoning on why the code needs to be changed, namely because it is not wrapped according to the project's standard, without explicitly giving a suggestion on how to fix it.
Examples of comments containing both an explanation and suggestion are shown in C5 and C6 of Table~\ref{tab:example_label_explanation}. 
In C5, the reviewer asked the author to change the code as the \textit{libchrome} library is being migrated. 
For C6, the reviewer asked the author to change the variable name as the variable that is used now no longer exists. 

\begin{table}
\caption{Examples of labeled code review comments. The extracted explanation parts are highlighted in the `Explanation' column, while the suggestion parts extracted are highlighted in the `Suggestion' column.}
    \label{tab:example_label_explanation}
    \centering
    \begin{tabular}{|p{0.03\linewidth} | p{0.3\linewidth} |p{0.3\linewidth} |p{0.3\linewidth} |}
    \hline
      \textbf{\#} &\textbf{Code Review Comment}  & \textbf{Explanation} & \textbf{Suggestion}                     \\  \hline
      C1 &\textbf{ASSERT}     & - &   \textbf{ASSERT}                   \\ \hline
        C2 & \textbf{These `DupCertBuffer().get()' could all be `GetCertBuffer()'}       & - &   \textbf{These `DupCertBuffer().get()' could  all be `GetCertBuffer()'}                \\ \hline
        C3 &\textit{this doesn't really say what gets recorded}     & \textit{this doesn't really say what gets recorded} &   -                   \\ \hline
        C4 &\textit{This isn’t wrapped according to our standard.}      & \textit{This isn’t wrapped according to our standard.}
          &   - 
        \\ \hline
         C5    & \textbf{Use `base::Value::Dict' if possible.}  \textit{libchrome is migrating from this usage to `base::Value::Dict'.} & \textit{libchrome is migrating from this usage to `base::Value::Dict'.}
         &   \textbf{Use `base::Value::Dict' if possible.}              \\ \hline
        C6 & \textbf{PowerButton}, \textit{since there is no power\_button symbol} & \textit{since there is no power\_button symbol}
         &   \textbf{PowerButton}                 \\ \hline
    \end{tabular}
\end{table}

\begin{figure}
\centering
\includegraphics[scale=0.42]{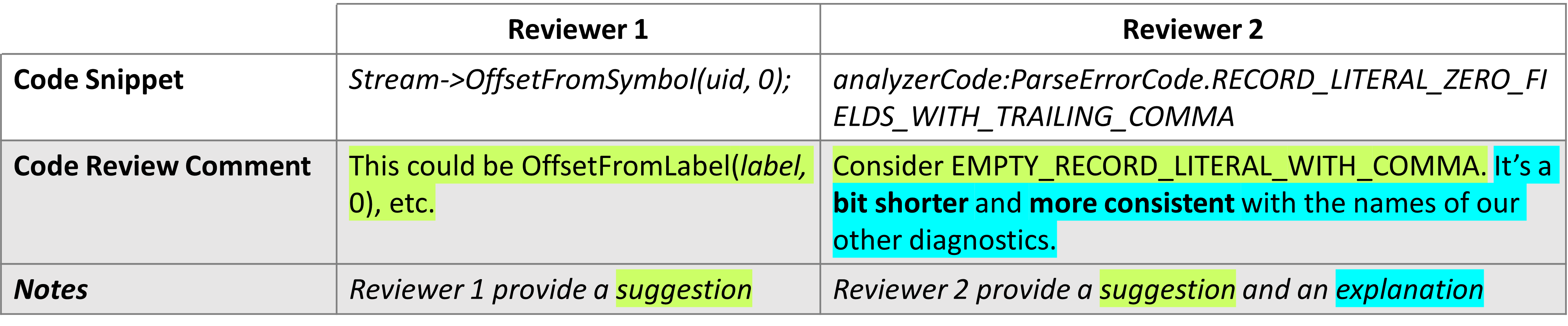}
\caption{Examples of code review comments in similar code issues.}
\label{fig:motivation_3}
\end{figure}

The results of our labeling are shown in Figure~\ref{fig:venn_rq1}. 
We observed that close to half (42\%) of the code review comments only had a suggestion. 
The ratio between code review comments that only have an explanation and those that only have a suggestion is 1:4. 
We also found that if the reviewer gave an explanation, it was usually accompanied by a suggestion in 79\% (297 out of 378) of the code review comments.
Suggestions provided in code review comments may carry a tacit or understood explanation. However, the interpretation of these implicit explanations can vary among developers. What appears obvious to one developer might not be as clear to another. For example, in  Figure~\ref{fig:motivation_1} the reviewer suggested removing part of the code without explanation, assuming it was self-evident, while the author did not understand the rationale behind this suggestion. 
Similarly, Figure~\ref{fig:motivation_3} showcases the simple task of updating a variable name. One reviewer omitted an explanation, assuming the suggestion was self-explanatory, whereas another provided reasoning for their suggestion. This variability highlights that what is obvious to one person may not be clear to another. By providing explicit explanations, we can bridge these gaps in understanding and improve the overall effectiveness of the code review process. 

The labeling results for the Android, Chromium, Dart, Gerrit, and Go projects are presented in Table~\ref{tab:dist_useful}. 
The analysis reveals that a noteworthy percentage of code review comments in all projects solely contain a suggestion, which aligns with findings on the whole dataset. Specifically, the minimum percentage of such cases is 36\% in the Dart project, while the maximum is 45\% in the Gerrit project.
Consistent with our previous discovery that explanations are typically accompanied by suggestions (79\% of cases), we observe that the percentages of these cases are 81\%, 74\%, 89\%, 67\%, and 87\% for the Android, Chromium, Dart, Gerrit, and Go projects, respectively. These findings reinforce the consistency across projects, supporting the notion that they share similar characteristics as revealed in the aggregated data.

\begin{tcolorbox}
{\textbf{RQ1 Findings:} 
42\% of the investigated code review comments only contained suggestions on how to improve or modify the code (without explanation). 
Furthermore, in the majority of cases (79\%) where an explanation was provided, the latter was accompanied by a suggestion.}
\end{tcolorbox}

\subsection{RQ2: What Types of Explanations Are Found in Code Review Comments?}
To address this RQ, we conducted open card sorting on the labeled data obtained from RQ1. 
For this, we randomly selected 298 code review comments (see Section~\ref{sec:rq2_method}) that contained explanations. 
Seven distinct categories referring to different types of explanations in code review comments emerged from our open card sorting.
In the following, we describe each category of explanations and give examples from our dataset.
The full dataset is available online~\cite{replication_package}:

\vspace{0.2cm}\noindent{\textbf{Category 1: By providing a rule or \underline{principle}.}}
Explanations of this category contained explicit rules, principles, or guidelines mentioned by the reviewer for the author to comply with.
These guidelines were presented in different forms for instance as a reference to a project guideline on writing code or as a reference to language standards.
For example, the explanation ``chromium style guide says \textit{you should not handle DCHECK() failures, even if failure would result in a crash.}  See [URL]''\footnote{\url{https://chromium-review.googlesource.com/c/chromium/src/+/3936275/20..24/chrome/browser/ash/login/quick_unlock/pin_backend.cc\#b576}} provided a reference to the internal Chromium coding guide for handling DCHECK() failures. 
Another example explanation is ``When the target name is the same as the directory name, we omit it.''\footnote{https://chromium-review.googlesource.com/c/chromium/src/+/3936363/2..3/ios/chrome/app/BUILD.gn\#b407}
In this example, the reviewer pointed to a seemingly implicit project rule regarding naming (i.e., naming convention).

\vspace{0.2cm}\noindent{\textbf{Category 2: By providing \underline{similar examples}.}} Explanations of this category contain examples from a code section or branch for the author to adjust the changes to match the reference. 
For example, in the review comment, ``There are couple of similar implementations. [url]''\footnote{\url{https://chromium-review.googlesource.com/c/chromium/src/+/3936522/2..5/third_party/blink/web_tests/external/wpt/common/dispatcher/remote-executor.html\#b14}}, the reviewer provided a link to a similar implementation from the other parts of the code. 
In another example, the reviewer commented: ``(As is) enum FittingType in constants.ts has a different case ordering.'',\footnote{\url{https://chromium-review.googlesource.com/c/chromium/src/+/3936487/4..9/chrome/browser/resources/pdf/pdf_viewer.ts\#b535}} highlighting that the case ordering in an enum should match the existing code.

\vspace{0.2cm}\noindent{\textbf{Category 3: By providing a \underline{test scenario} or a test condition.}}
This type of explanation contains a scenario or condition that should be tested or considered when writing the code.
It presents a ``what if'' scenario following which some factors are different along with the result of changing those factors. 
Described scenarios may have different severity levels but most reviewers are concerned with program failure scenarios.
For instance, in this review comment, ``There's not really point of EXPECT\_TRUE here - if it's false the test will crash.'',\footnote{\url{https://chromium-review.googlesource.com/c/chromium/src/+/3936665/3..4/chrome/browser/chromeos/extensions/desk_api/desk_api_extension_manager_unittest.cc\#b73}} the reviewer highlights that the test will crash if the value is False. 
In another example, the reviewer uttered: ``ASSERT (given that the test crashes if manifest\_value is null)''\footnote{\url{https://chromium-review.googlesource.com/c/chromium/src/+/3936665/3..4/chrome/browser/chromeos/extensions/desk_api/desk_api_extension_manager_unittest.cc\#b250}} pointing out that the test will crash if the condition is not met.

\vspace{0.2cm}\noindent{\textbf{Category 4: By discussing \underline{future implications}.}}
Explanations of this category mention something that may happen in the future that would affect the current change. 
The difference with the previous category is that comments of this category do not necessarily give the change of input that can make the code fail, it explains something that may happen in the future to make the developer need to change the affected code section again. 
For example, we sorted the comment ``I think there are other features in the future that may freeze to; it just wasn't initially clear to me that this was related to page freezing.''\footnote{\url{https://chromium-review.googlesource.com/c/chromium/src/+/3936590/5..6/third_party/blink/public/platform/web_media_player.h\#b184}} under "future implication" rather than "by providing a scenario or condition to test" since this explanation considers the possibility of additional features in the future that may also require page freezing.

\vspace{0.2cm}\noindent\textbf{Category 5: By expressing \underline{personal preferences} or opinion regarding the reviewed code.}
In review comments of this category, reviewers express a personal preference or opinion about the code that is being reviewed (i.e., code written by the author).
For example, in the code review comment: ``\textit{\_Prepare} is a little bit confusing for me.'',\footnote{\url{https://chromium-review.googlesource.com/c/chromiumos/platform/factory/+/3936133/1..3/py/tools/finalize_bundle_unittest.py\#b179}} the reviewer expressed the confusion relating to a specific function name `\_Prepare' that was written by the author.

\vspace{0.2cm}\noindent{\textbf{Category 6: By stating an \underline{issue} with the code under review.}}
In this category, the code reviewers point out facts about the current status of the code or possible 
root causes of an issue. 
For example, in this code review comment: ``libchrome is migrating from this usage to \textit{base::Value::Dict}.'',\footnote{\url{https://chromium-review.googlesource.com/c/chromiumos/platform2/+/3935924/1..4/runtime_probe/functions/mipi_camera.cc\#b318}} the reviewer points to a migration of the libchrome library, which has a potential impact on the code change.
In another example, the reviewer states: ``you don't need the time zone, since you are not touching it.''\footnote{\url{https://chromium-review.googlesource.com/c/chromium/src/+/3936353/4..6/chrome/browser/ui/android/omnibox/java/src/org/chromium/chrome/browser/omnibox/suggestions/answer/AnswerSuggestionProcessorUnitTest.java\#b84}} recognising an unnecessary component usage (timezone) that is causing the issue.

\vspace{0.2cm}\noindent{\textbf{Category 7: By providing the \underline{benefit} of applying the reviewer's suggestion.}}
In this category, the code reviewers proposed a code change (i.e., suggestion) that the code author needs to apply and substantiated their suggestion with a statement highlighting the benefit of applying the suggested code change, for example, with regards to performance (\textit{faster}) or code readability (\textit{more readable}), etc. This focuses on the suggestion given by the reviewer as opposed to Category 5, which concentrates on the code written by the author.   An example code review comment reads ``I think \textit{ExpectContainsAuthIntent} is a little clearer'',\footnote{\url{https://chromium-review.googlesource.com/c/chromiumos/platform/tast-tests/+/3936642/1..5/src/chromiumos/tast/common/cryptohome/helpers.go\#b31}} highlighting that the suggestion, i.e. changing the current function name ``ExpectHasAuthIntent'' to ``ExpectContainsAuthIntent'' adds clarity to the current code.
In another example, from this code review comment, ``If so can you mention where these reside in a comment to help authors keep them in sync?''\footnote{\url{https://chromium-review.googlesource.com/c/chromium/src/+/3936491/5..7/chrome/browser/resources/settings/performance_page/performance_metrics_proxy.ts\#b4}} the reviewer highlights that the proposed suggestion is intended to help the authors keep the comment in sync.

\begin{figure*}
\centering
\includegraphics[scale=0.52]{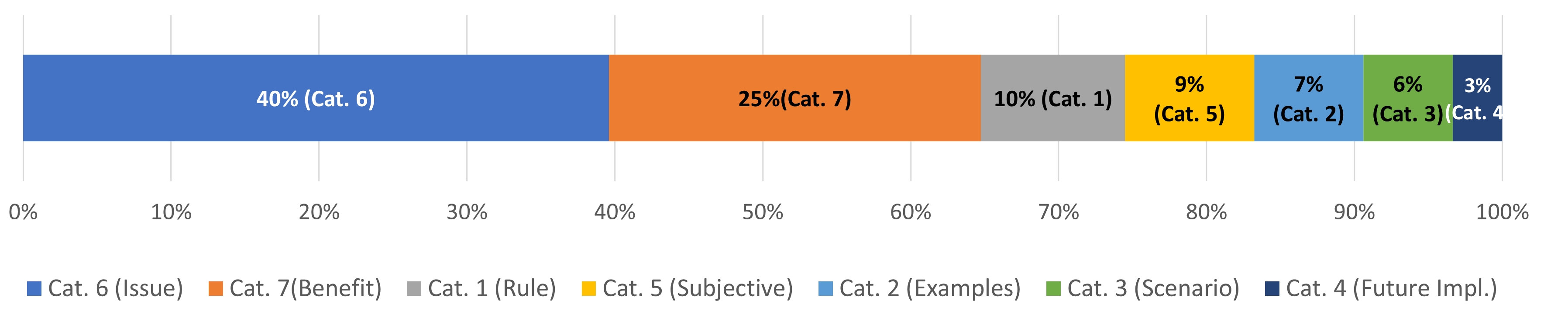}
\caption{Distribution of explanation types in the manually analyzed code review comments (N=298).}
\label{fig:distribution_rq2}
\end{figure*}

Figure~\ref{fig:distribution_rq2} shows the distribution of the code review comment explanation types from our dataset. 
We found that \textit{Category 6}, which states the current situation of the code or gives the root cause of why the current code needs improvement, is the category that reviewers predominantly use (40\%). 
Second to that is \textit{Category 7} (25\%), in which reviewers provide supporting statements for the suggestion they propose. 
\textit{Category 1} comes next in the ranking (10\%) and describes comments in which the reviewer points to rules or principles. 
These findings are consistent across all projects, with \textit{Category 6} and \textit{Category 7} occupying the top two positions, except for the Dart project, where the position is swapped.
These results suggest that reviewers tend to give a direct explanation about the root cause compared to other explanations such as failed scenarios and future implications.

\begin{tcolorbox}
{\textbf{RQ2 Findings:} 
We identified seven types of explanations in our data set. 
The explanation used most by reviewers (40\% of cases) states an issue with the code. The one used least refers to future implications (3\% of cases).
}
\end{tcolorbox}

\subsection{RQ3: How Does Reviewers’ Experience Affect the Frequency and Type of Explanations?  }
\subsubsection{Reviewer Experience and the Frequency of Explanation Given in Code Review Comments}
\begin{figure}
\centering
\includegraphics[scale=0.45]{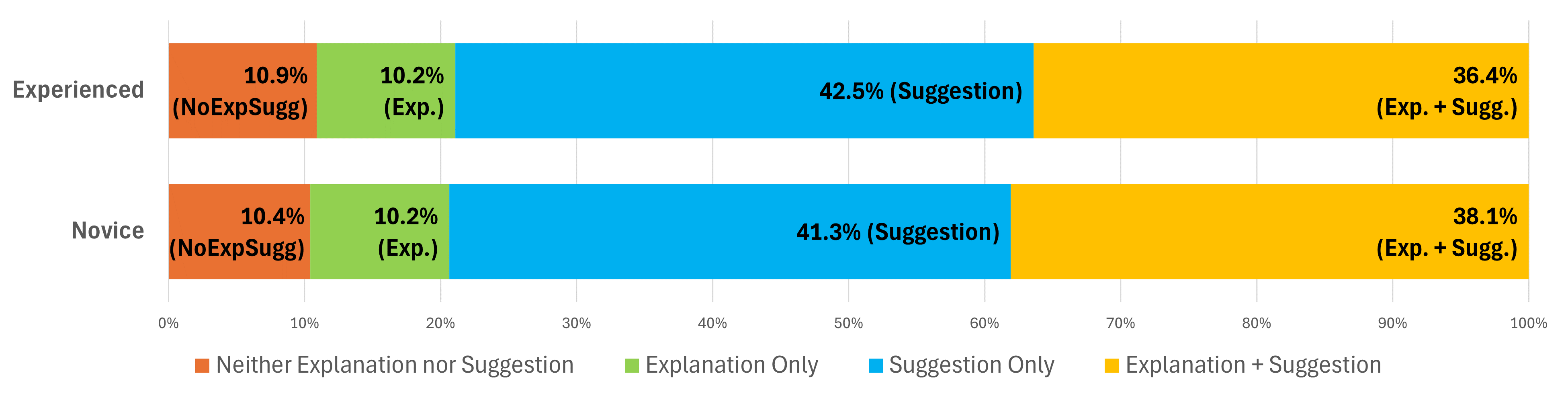}
\vspace{-0.3cm}
\caption{Distribution of explanations and suggestions among novice and experienced reviewers.}
\label{fig:distribution_rq3_1}
\vspace{-0.3cm}
\end{figure}

We re-investigated the labeled code review comments from RQ1, where we checked whether the comment contained an explanation only, a suggestion only, or both an explanation and suggestion, and this time we took into account the experience of the reviewer. Figure~\ref{fig:distribution_rq3_1} shows the distribution of the explanations in code review comments based on the experience level. For the frequency of explanations alone, both experienced and novice reviewers have the same percentage at 10.2\%. Meanwhile, the difference between novice and experienced reviewers in suggestion and explanation + suggestion are minimal, at 1.4\%, and 1.7\%, respectively. These differences were found to be statistically insignificant (p=0.9) when tested using the Chi-Square statistical test~\cite{tallarida1987chi} following previous studies~\cite{Kamienski2021, Taibi2017,Kosti2014,David2008}. These findings indicate a consistent distribution across both groups, suggesting there are no significant differences in the frequency of providing explanations. Both experienced and novice reviewers prioritize sole suggestions over sole explanations. When explanations are provided, they usually include suggestions, highlighting the importance of suggestions from the reviewers' perspectives, regardless of their experience level.

\subsubsection{Reviewer Experience and the Type of Explanation Given in Code Review Comments}
\begin{figure}
\includegraphics[scale=0.47]{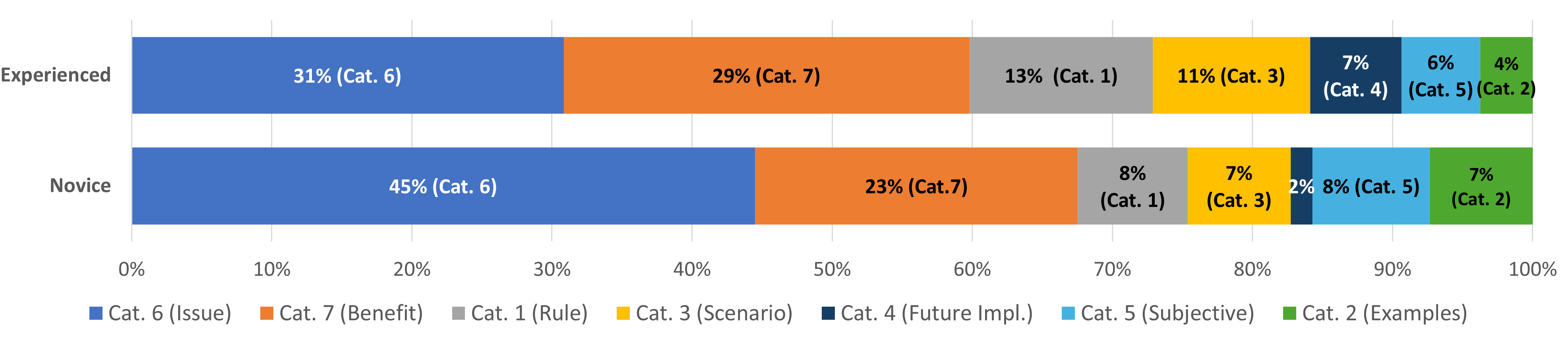}
\vspace{-0.3cm}
\caption{Distribution of explanation types among novice and experienced reviewers.}
\label{fig:distribution_rq3_2}
\end{figure}

Figure~\ref{fig:distribution_rq3_2} shows the distribution of explanation type between experienced and novice reviewers. The results show that the most common type of explanation used by both novice and experienced reviewers is Category 6, where the reviewer states the cause of the issue. Notably, novice reviewers have more code review comments in Category 6 compared to experienced reviewers, with a difference of 14\%. The second most frequent type of explanation for both novice and experienced reviewers is Category 7, where the reviewer explains the benefit of applying their proposed code change. Experienced reviewers use Category 7 more often than novices, with a 6\% difference.

In third place, experienced reviewers utilize Category 1, which involves providing a rule or principle, accounting for 13\% of their reviews. In contrast, novice reviewers are equally likely to use Category 1 and Category 5—the latter involving subjective feedback about the author's code. Experienced reviewers place Category 3 (providing a scenario) and Category 4 (explaining future implications necessitating code changes)  in the fourth and fifth positions, respectively. This differs from novice reviewers, who have Categories 3 and 4 as the last and second to last. 

The differences in the ranking of explanation categories between experienced and novice reviewers are more pronounced in the less common categories, which require more complex explanations compared to the more straightforward Categories 6 and 7. For instance, experienced reviewers have Category 4, a more challenging category that involves discussing future implications, in fifth place, whereas it falls to seventh place among novices. Similarly, Category 3, which requires providing a scenario, is ranked fourth in experienced reviewers but only sixth in novices. 

Overall, experienced reviewers demonstrate greater variation in their explanations beyond Category 6. 
The differences in usage percentages for Categories 1, 2, 3, 4, 5, and 7 between experienced and novice reviewers are 6\%, 5\%, -3\%, 4\%, -2\%, and 6\%, respectively. This suggests experienced reviewers have a broader range of explanatory skills.

Additionally, a Chi-Square statistical test (p=0.02) confirms statistically significant differences in the distribution of explanation types between experienced and novice reviewers. This finding underscores the impact of experience on the choice of explanations provided by reviewers. Novices tend to favor more straightforward explanations, such as identifying the root cause of an issue, while experienced reviewers are more likely to use a diverse array of explanation types. Particularly, experienced reviewers are more likely to discuss future implications compared to novice reviewers.  

\begin{tcolorbox}
{\textbf{RQ3 Findings:} 
We observed statistically significant differences in the types of explanations provided during code reviews between novice and experienced reviewers. Experienced reviewers demonstrate a broader range of explanation categories, indicating a greater diversity in their approach to code reviews compared to novices.  }
\end{tcolorbox}

\subsection{RQ4: To What Extent Can ChatGPT Generate a Specific Type of Explanation?}

\subsubsection{Prompt Engineering and Initial Evaluation}
To answer this RQ, we first created a targeted prompt for every explanation type. 
The prompts are created based on the explanation pattern and the category name. 
We then used these prompts in ChatGPT-July-6-2023~\cite{gptversion} version (which uses the GPT3.5 model). 
As discussed in Section~\ref{sec:motivation} and Section~\ref{sec:rq3_method}, a primary use case is to allow the authors of the reviewed code to select their preferred explanation and potentially reduce the need for further clarifications. 
One other use case is to support novice reviewers in providing various types of explanations. 
A prompt using all available information types (commit message, code line, code review comment, and code snippet) is shown in Figure~\ref{fig:prompt_transform}. To generate specific types of explanation, we need to provide \textit{[TARGET CATEGORY]} in the prompt which is the name of the explanation category as presented in Section 4.2. 
For instance, for Category 1, we have "\textit{Re-write the code review comment provided above so it includes an explanation that provides a \underline{rule or principle.}}"

\begin{figure}[hbtp]
    \begin{lstlisting}[frame=single, aboveskip=-0.0 \baselineskip,]
You are a code reviewer. Your task is to provide a different type of explanation for the highlighted code. Here is an example of a code review: 

Commit Message: [COMMIT MESSAGE]
Line [LINE NUMBER]: [LINE HIGHLIGHTED CODE] 
Code Review Comment: [CURRENT REVIEW] 
Code: [CODE SNIPPET]

Re-write the code review comment provided above so it includes an explanation that provides a [TARGET CATEGORY].
    \end{lstlisting}
    \caption{Prompt to generate a specific type of explanation using all information considered.}
    \label{fig:prompt_transform}
\end{figure}

Furthermore, we proposed different variations of the prompts based on the information input (see Section~\ref{sec:rq3_method}).
In total, we have 4 different variations of the prompt which are "Base" (includes only the base inputs), "Base+Line" (includes both the base and line inputs), "Base+Commit" (includes both the base and commit message inputs), and "All" (includes the base and all additional inputs).
We utilized 5 different code review comments that come from different categories. 
Then, we generated a specific type of explanation, for each prompt we got 30 code review comments. 
In total, we evaluated 120 code review comments that came from different prompts.

The results of the first evaluation are presented in Table~\ref{tab:RQ3_eval1}. 
We observed that the "Base+Line" prompt yielded the highest number of correct type explanations, while the "All" prompt obtained the highest number of correct explanations overall. Conversely, the baseline prompt "Base", which utilized no additional information, received the lowest scores for both type and explanation correctness. Considering the overlapped cases (i.e., instances where both the type and explanation are correct), the results remain consistent, with prompts that included additional information outperforming the ones without it.  
This emphasizes the value of the additional information in improving the generated explanations.

The "Base+Line" prompt achieved the highest total and overlapped score for type and explanation correctness, surpassing both the "All" and "Base+Commit" prompts. 
This difference may arise from the fact that not every commit message is directly relevant to the code review comment. 
For instance, consider this commit message: ``\textit{runtime\_probe: Add mipi\_camera probe function. 
Integrate cros-camera-tool into runtime probe to implement probe function for MIPI cameras,}" while the code review comment states: ``\textit{Use base::Value::Dict if possible. libchrome is migrating from this usage to base::Value::Dict.}" 
The commit message is unrelated to the review comment, as it addresses a different aspect of the codebase. 
Including this irrelevant information in the prompt may hinder ChatGPT's ability to generate accurate comments. 
Conversely, the line number information is consistently related to the specific code, which can aid ChatGPT in generating targeted comments.

\begin{table}[]

\caption{Number of code review comments that have the correct type and correct explanation.}
\vspace{-0.2cm}
\label{tab:RQ3_eval1}
\setlength{\tabcolsep}{0.8em}
\begin{threeparttable}
\begin{tabular}{|l|r|r|r|r|}
\hline
\multicolumn{1}{|c|}{\textbf{Prompt}\tnote{\S}} & \multicolumn{1}{c|}{\textbf{Correct Type}} & \multicolumn{1}{c|}{\textbf{Correct Expl.}} & \multicolumn{1}{c|}{\textbf{Total}\tnote{\textdagger}} & \multicolumn{1}{c|}{\textbf{Overlapped}\tnote{\textdaggerdbl}} \\ \hline
Base                               & 23                                             & 24                                                    & 47 & 19                                  \\ \hline
Base+Commit                              & 26                                             & 26                                                    & 52 & 23                                 \\ \hline
Base+Line                            & 27                                             & 27                                                    & 54         & 24                         \\ \hline
All                                   & 24                                             & 28                                                    & 52    & 23                              \\ \hline
\end{tabular}
\begin{tablenotes}
 \footnotesize
 \item[\S] Name of the four prompts variants. \textdagger Total is the sum of the correct type and correct explanation. 
 \item[\textdaggerdbl] Overlapped is the number of code review comments that have both correct type and correct explanation.
    \end{tablenotes}
\end{threeparttable}
\vspace{-0.2cm}
\end{table}

\subsubsection{Extended Evaluation of Best-Performing Prompt}
For further evaluation, we utilize the "Base+Line" prompt, which includes the base information along with the highlighted line, to generate code review explanations. We chose this prompt as it provided the best results in the previous step.

In this subsequent evaluation, we evaluated a total of 90 code review comments, with 15 reviews from each category. 
The results from the evaluators revealed encouraging scores for both informativeness and clarity metrics, averaging at 6.13 and 6.07, respectively. The slim margin of 0.06 between the average informativeness and clarity scores highlights an effective balance achieved by ChatGPT in providing clear and informative explanations. These results show that the explanations generated by ChatGPT tend to be both clear and informative.

Meanwhile, with respect to the correctness metrics, the average shows 88 and 89 out of 90 (96\% and 98\%) have the correct explanation category and correct explanation. These results highlight ChatGPT's ability to generate specific types of explanations accurately, capturing both the intended category and the content of the explanation itself. Take into account this example of code review comment written by a human reviewer: `\textit{`Nit: maybe call this `tmp' to follow the naming convention."}, which is transformed by ChatGPT to \textit{Category 6} (i.e., required providing the issue with the code under review): \textit{``Nit: maybe call this `tmp' to follow naming convention. The current variable name `TMP' deviates from the established naming convention, which typically uses lowercase letters for variable names''}.
The generated explanation by ChatGPT demonstrates the correct intended category and highlights that the current variable deviates from the naming convention. Meanwhile, comparing the code review comment written by a human reviewer and the code review comment generated by ChatGPT, ChatGPT still provides the correct issue. The supplementary information provided by ChatGPT, stating that the current variable deviates from the naming convention, is also correct.

Despite the favorable overall evaluation results, there were instances where evaluators found the generated explanations inconsistent with the assigned categories or lacking in correctness. 
One such example involved a code review comment written by a human reviewer stating: \textit{``Nit: maybe call this `tmp' to follow the naming convention.''}, this was transformed into a code review comment by ChatGPT as: \textit{``Nit: It would be better to follow the naming convention and call this variable `tmp' instead of `TMP'. Consistency in variable naming helps improve code readability and maintainability. For example, if we have another variable named `tmp2' in the future, it would be easier to understand their purpose and differentiate between them.''} The intended category for this transformed code review comment was \textit{Category 2}, which required providing similar examples. 
However, the ChatGPT-generated comment focuses on suggesting a variable name change for better differentiation, rather than offering examples of variables following the same naming convention. On further analysis,  we observed that \textit{Category 2} had the highest number of generated code review comments labeled as incorrect. 
This shows that ChatGPT still has a limitation on generating similar examples from the input code review comment and the code snippet.

Regarding the correctness of the explanation, ChatGPT highlighted that \textit{tmp} is easier to differentiate from \textit{tmp2} instead of \textit{TMP}. However, the evaluators disagreed. There was no significant difference between distinguishing \textit{tmp} from \textit{tmp2} and \textit{TMP} from \textit{tmp2}, and the suggested change did not lead to improved understanding or differentiation. 
The highest number of generated code review comments that have incorrect explanations comes from \textit{Category 2} 
which shows that there is still a limitation on generating accurate examples by ChatGPT. Consequently, reviewers are advised to exercise caution when relying solely on ChatGPT-generated explanations, especially in critical review comments related to bugs or potential code-breaking issues.

As a first step toward explanation generation, we have developed a custom version of ChatGPT to help developers efficiently transform explanations,  allowing them to select from a predefined list of explanation categories. The tool also provides guidance on the information required as input. As illustrated in  Figure~\ref{fig:gpt_custom_0}, users can select a predefined question to obtain more detailed information about the prompt. Figure~\ref{fig:gpt_custom_1} shows the various explanation options that ChatGPT can generate, along with the specific input needed from the users. Figure~\ref{fig:gpt_custom_2} shows an example from the custom version of ChatGPT generating several types of explanations chosen by the developers. The prompts used are inline with those discussed in this paper. The tool can be accessed at: 
\url{https://chatgpt.com/g/g-ymkMdlDhx-explanation-code-review-transform.}

\begin{figure} 
\centering
\begin{subfigure}[b]{.49\linewidth}    \includegraphics[width=\linewidth]{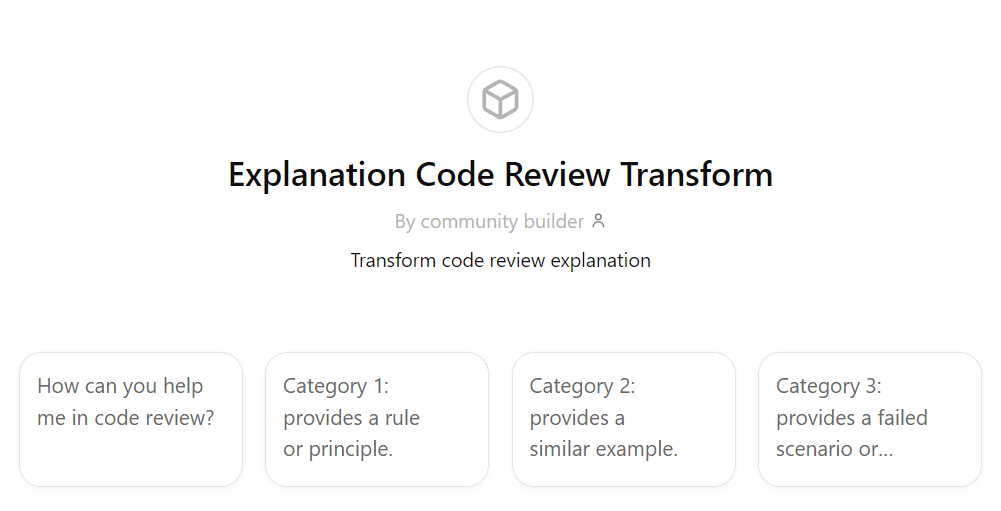}
\caption{Sample questions.}\label{fig:gpt_custom_0}
\end{subfigure}

\begin{subfigure}[b]{0.49\linewidth}
\includegraphics[width=\linewidth]{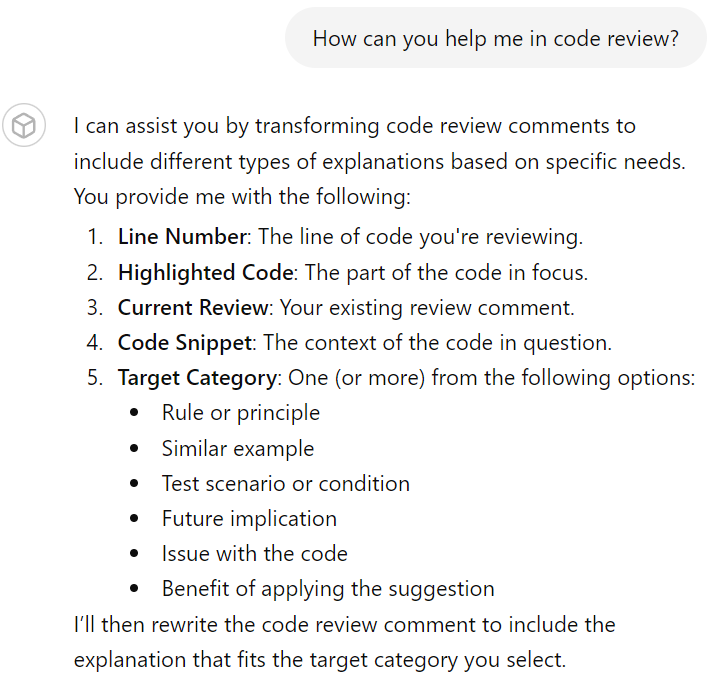}
\caption{Sample answer.}\label{fig:gpt_custom_1}
\end{subfigure}        
\begin{subfigure}[b]{0.49\linewidth} \includegraphics[width=\linewidth]{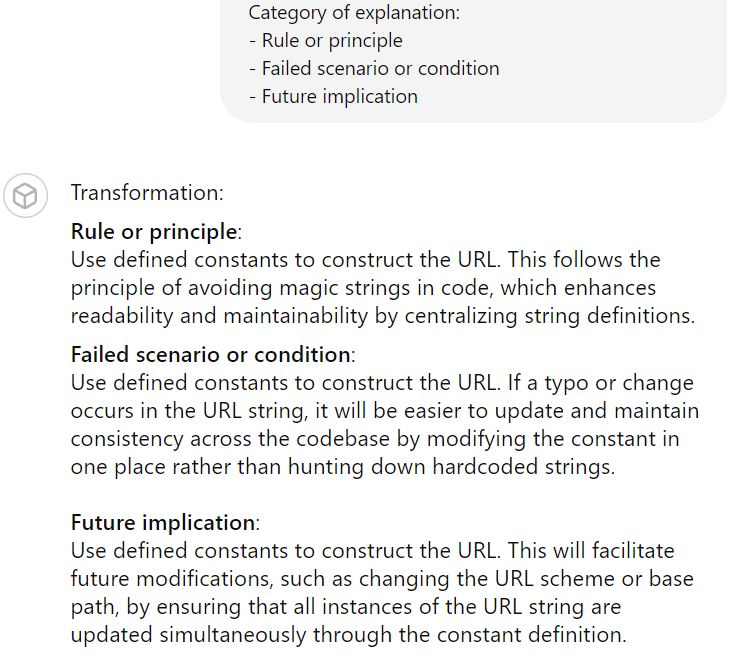}
\caption{Sample transformation.}\label{fig:gpt_custom_2}
\end{subfigure}
\caption{Custom version of ChatGPT to transform explanation in code review comments.}
\label{fig:gpt_custom}
\end{figure}

\begin{tcolorbox}
{\textbf{RQ4 Findings:} 
Out of 90 cases, ChatGPT-transformed explanations matched the specified type in 88 instances and provided correct explanations in 89 cases. Overall, ChatGPT exhibits robust results in terms of clarity and informativeness, scoring 6.13 and 6.07 on a 7-level Likert scale, respectively. } 
\end{tcolorbox}

\section{Discussion}
\label{sec:discussion}

\subsection{Implications for Practice}
\subsubsection{Effective Review Explanations Are Important but Require Time and Skills} 
We hope that our study will increase awareness about the importance of explanations in code reviews.
Our results reveal that a substantial 42\% of the analyzed code reviews contain only suggestions without explanations. 
We think that this is a rather low ratio, suggesting that reviewers should, in general, articulate the reasons behind their recommendations more frequently. 
By providing explanations, the reviewers can help the authors not only understand \textit{what} needs to be changed but also \textit{why}. 
This can also lead to more productive and collaborative review processes, as authors are more likely to accept and act on feedback when they grasp the rationale or argue why their versions are better.
One potential reason for the lacking explanations is the reviewers may lack time or priority for it. 
While this can be mitigated in practice by better awareness, guidelines,  and tool support, future studies might focus on quantifying the impact of explanation. 
Another potential reason is the lack of skills. 
In fact, our findings show that experienced reviewers use a broader variety of explanations suggesting that experience and skills likely play a role for effective explanation.

\subsubsection{LLMs Can Help Reviewers Generate Alternative Explanations}
Providing suitable explanations requires effort from reviewers. To reduce this, reviewers could use LLMs to generate alternative explanations. 
By using the seven explanation categories identified in our work, reviewers can reason and structure their feedback more efficiently, making it easier for authors to understand their suggestions and intentions. 
For instance, if a code author continues to have questions after the initial explanation, a reviewer might use an LLM to generate an alternative explanation. 
The various explanation types serve as different “templates” for the LLM to create review comments depending on the reviewers' needs. 
Additionally, in RQ3 we found that experienced reviewers tend to give more diversified explanations compared to novices. 
This suggests that LLMs could also help novices in creating other types of explanations.
Our prompts and the tool discussed in Section \ref{sec:results} can serve as a first step to this end.

\subsubsection{Authors Can Proactively Seek Explanations and Eventually Be Assisted by LLMs Too} 
Through a proactive approach, authors can gain a better and quicker understanding of not only what aspects require improvement but also the rationale behind these suggested enhancements in a code review. 
This empowers authors to make well-informed decisions and, e.g., implement modifications that align with the project goals and standards. 
Code authors might specify the kind of explanation they miss, such as a similar example they can look at. This can lead to need-driven knowledge sharing and to improvement of the communication effectiveness among developers.

In particular, authors can themselves use LLMs to tailor the type of explanations needed.
Authors can utilize this method to get tailored explanations, potentially aiding a more nuanced understanding of the feedback. 
Our evaluation results of ChatGPT in transforming explanations show the potential for integrating such an LLM-driven approach into the code review workflow. 
Instead of waiting for or getting a single explanation, authors could generate and explore multiple explanations. 
This can minimize back-and-forth communication between authors and reviewers, which in turn would speed up the review process, particularly in continuous deployment settings.
In the broader context, this can contribute to better on-boarding of new contributors \cite{Stanik:ICSME:2018} and to reducing pull requests abandonment \cite{khatoonabadi2023wasted}.
In fact, Khatoonabadi et al.~\cite{khatoonabadi2023wasted} conducted a mixed-method study of 4,450 abandoned pull requests and found that lengthy reviews are a contributing factor leading authors to stop working on a change.
When analyzing the reasons behind abandonment, the researchers found that the leading cause is that the authors had difficulty addressing maintainers' comments.

\subsection{Implications for Future Research}

\subsubsection{Understanding Explanation Needs and Styles as Part of Knowledge Sharing}

Our work only provides starting evidence about the prevalence and typology of explanations in code reviews. 
Future researchers should investigate how various explanation types and “explanation styles” in code reviews affect the review outcome and the overall review process. 
For instance, it remains unclear how specific types of explanations, their timing, or order impact code quality, developer engagement, and the efficiency of collaboration. 
Answering these questions likely provides insights for improving code review practices.
Moreover, it is also worth investigating what influences the explanation needs~\cite{bouraffa2020two} (and preferences) of both the reviewers and authors. 
We took a first step in this direction analyzing the difference between novices and experienced reviewers. 
Gathering additional data about the reviewers or using different research methods (as controlled experiment or interviews) can provide a more thorough understanding towards a personalized, need-driven code review explanations and knowledge sharing between developers in general \cite{Maalej:BotSE:2023}.

Generally, explanations (i.e.~rationale or answers to \textit{why} questions) represent a valuable type of development knowledge that often remains in the head of developers (tacit knowledge) ~\cite{maalej2013patterns, maalej2014comprehension}. 
Our work can help identify---and eventually externalize---this knowledge from code reviews for reusing it in other development tasks~\cite{happel2008potentials,Maalej:BotSE:2023}. 
For instance, explanations mentioning coding rules and practices to follow or test cases to cover could potentially be useful for new project members \cite{Stanik:ICSME:2018} and other developers to extract and share in the documentation. 
Since manually writing informative review comments is time-intensive, future researchers should focus on developing automated solutions to generate specific explanation types directly from the code context.

\subsubsection{What Makes a Good Explanation?}
Our work stops at the reviewer's perspective and does not investigate whether and when explanations are actually effective---that is whether a certain explanation is of a high or low quality. 
While researchers have investigated the quality of software documentation in the past\cite{maalej2013patterns, Tang:MSR:2023}, explanations in review comments are rather informal and different. Therefore, it remains unclear what makes a good explanation.

One potential influencing factor is the preciseness and quality of the language used. 
Another is the combination of explanations and arguments---an aspect this work does not address.
Our categorization was based on the entire review comment. 
When analyzing explanations, we observed that in some cases several explanation categories are combined in one comment. Assuming that multi-typed explanations would likely reside in lengthier comments, we also sampled comments that have a text length above the 75th percentile. 
Out of the 71 long code review comments, we found only nine that have multiple types of explanations. 
For example:
“\textit{We have another memory leak here. This code should be replaced with: `int sizes = features.size(); TfLiteInterpreterResizeInputTensor(interpreter, 0, \&sizes, 1); TfLiteInterpreterAllocateTensors(interpreter);` That way we don't need to free the memory as it will be allocated on the stack and freed when the function exits.}”\footnote{\url{https://android-review.googlesource.com/c/platform/frameworks/opt/gamesdk/+/2238874/10..11/src/memory\_advice/core/predictor.cpp\#b92}}
In this comment, the reviewer first point to a problem—a memory leak (Category 6). The reviewer then proposes a specific code change, followed by an explanation of the benefits (Category 7). 
However, when considering the entire code review comment, the primary focus is on identifying the issue (i.e.~the memory leak). 
The subsequent sentences, including the suggestions and its benefits, serve to reinforce the resolution of the initially stated issue. 
Thus, in RQ1, this entire comment is categorized under Category 6. 
We included the nine cases with multiple types of explanations in our replication package. 
This observation opens up the possibility of combining various types of explanations to provide more comprehensive--possibly more useful---feedback. 
Future studies can investigate when an explanation should be short or more detailed and when it should be single-typed or multi-typed.
Strategies of how to effectively combine explanations into single review comments are yet to be investigated.

\subsection{Threats to Validity}
\label{sec:threats}
\subsubsection{Internal Validity}
A potential threat to internal validity is our selection approach for useful code reviews.
For this, we followed the previous work by Bosu et al.~\cite{bosu2015characteristics}. We also adopted the method proposed by Zhang et al.~\cite{zhang2020sentiment} as an alternative to the original internal sentiment analysis tool, which was not accessible at the time of this research.
While this adaptation may introduce additional noise, we believe this to be minimal. 
Zhang et al.~\cite{zhang2020sentiment} demonstrated through extensive evaluation that their method outperforms 5 existing approaches across 6 datasets.
We make our implementation of the decision tree as well as our script to retrieve Gerrit code reviews public~\cite{replication_package}, to enable other researchers to validate our findings.

Another internal validity concern is a potential mislabeling in our data. To ensure the quality of the labeling, we manually analyzed the data over multiple iterations. 
We revisited categorizations several times over a year—in December 2022, April 2023, August 2023, and June 2024.
However, it remains possible that we mislabeled some examples. We also did not report the inter-rater agreement for the explanation categories. We instead employed a collaborative approach (open card sorting) that allowed us to discuss and resolve any differences immediately during the categorization process.  As
discrepancies were inherently resolved through discussion until a unanimous decision was reached, we cannot calculate the inter-rater agreement score. 
We included our labeled code review dataset as part of the replication package so other researchers can validate our findings.

\subsubsection{External Validity}
The generalizability of our findings is subject to potential threats too. 
One potential risk arises from the scope of our evaluations on the ability of ChatGPT to generate specific types of explanations. 
In the first phase of RQ4 analysis in our study, we assessed 120 samples for accuracy and correctness. This was followed by an evaluation of 90 samples in the second phase (generated using the best-performing prompt from phase one) to assess correctness, type correctness, clarity, and usefulness. The combined efforts for both phases total approximately 115 hours. Should we attempt to extend these evaluations to encompass all generated samples, the first phase alone would require an estimated 1,192 hours per evaluator. This figure is derived from the assessment of 298 original categories, each expanded through transformations into five additional categories and evaluated across four different prompts, resulting in a total of 7,152 samples, with each requiring about 10 minutes for assessment. Due to these constraints, previous studies also have often utilized relatively small sample sizes for similar evaluations~\cite{geng2024large, chen2020stay, kang2024human}. For instance, Geng et al.~\cite{geng2024large} evaluated 100 samples in their study.

Furthermore, we focused our analysis on all Google projects available on Gerrit. We selected Google projects because Google's code review guidelines specifically highlight the necessity of explanation in code review comments. Although we studied multiple projects, the results may not be generalizable to projects using other code review platforms. 
Furthermore, to maintain high quality and relevance in our manual analysis, we filtered out only the useful code review comments. However, it is important to note that while some reviews may not have been deemed useful through automatic filtering, they may still contain valuable insights and explanations.

\subsubsection{Construct Validity}
A potential threat to construct validity arises from the variability in ChatGPT-generated explanations caused by different prompts.
To create our prompts, we analyzed patterns from different types of explanations and followed the approaches from previous studies~\cite{zhou2022large,xia2023keep} that used prompts for ChatGPT. 
While different prompts and models may produce different explanations, we think that the overall trends and findings of our work remain valid.

\section{Conclusion and Future Direction}
\label{sec:conclusion}
Code review is a central activity in software development.
While numerous studies have attempted to automate code review activities, there has not been much work on understanding how reviewers actually bring their points across by explaining their feedback to authors.
This study aims to fill this gap by taking a close look at reviewer feedback in code review comments.

We collected a useful code review dataset from Gerrit and analyzed how often reviewers used explanations in their first review comments. 
We found that many of the comments in our data (i.e., 42\%) only contain improvement suggestions without explanations. 
At the same time, the majority (79\%) of review explanations are accompanied by a suggestion. 
When manually analyzing these explanations we identified seven distinct categories. 
We found that reviewers typically use Category 6, where they state an issue with the code under review. 
When correlating developers' experience with the explanations, we found that experienced reviewers are more likely to use diverse categories and to discuss potential future implications when explaining their review suggestions.
We also evaluated how well can ChatGPT transform different types of explanations using different prompts. 
Our results show that in 88 and 89 out of 90 cases, ChatGPT successfully generated a correct explanation and an explanation of the intended type, respectively. Moreover, the generated explanations demonstrated a high level of informativeness and clarity.

For future work, we plan to leverage LLMs not only to suggest different types of explanations but also the full code review comment (which includes the identification of the target location and type of explanation needed). 
We also plan to explore scenarios where an explanation is not initially provided during a code review, and the developer subsequently requests clarification. 
Specifically, we intend to investigate the effectiveness of prompting ChatGPT directly with the developer's question to generate the required explanation. 
Finally, substantial empirical work is needed to gather additional insights from developers (e.g. through interviews and experiments) regarding their explanation needs, styles, and investigate which explanations are most useful in which context.


\bibliographystyle{ACM-Reference-Format}
\bibliography{main}


\begin{thebibliography}{85}


\ifx \showCODEN    \undefined \def \showCODEN     #1{\unskip}     \fi
\ifx \showDOI      \undefined \def \showDOI       #1{#1}\fi
\ifx \showISBNx    \undefined \def \showISBNx     #1{\unskip}     \fi
\ifx \showISBNxiii \undefined \def \showISBNxiii  #1{\unskip}     \fi
\ifx \showISSN     \undefined \def \showISSN      #1{\unskip}     \fi
\ifx \showLCCN     \undefined \def \showLCCN      #1{\unskip}     \fi
\ifx \shownote     \undefined \def \shownote      #1{#1}          \fi
\ifx \showarticletitle \undefined \def \showarticletitle #1{#1}   \fi
\ifx \showURL      \undefined \def \showURL       {\relax}        \fi
\providecommand\bibfield[2]{#2}
\providecommand\bibinfo[2]{#2}
\providecommand\natexlab[1]{#1}
\providecommand\showeprint[2][]{arXiv:#2}

\bibitem[AI(2024a)]%
        {Meta_LLaMA3}
\bibfield{author}{\bibinfo{person}{Meta AI}.} \bibinfo{year}{2024}\natexlab{a}.
\newblock \bibinfo{title}{Meta LLaMA}.
\newblock \bibinfo{howpublished}{\url{https://ai.meta.com/blog/meta-llama-3/}}.
\newblock


\bibitem[AI(2024b)]%
        {Mistral_Mixtral}
\bibfield{author}{\bibinfo{person}{Mistral AI}.} \bibinfo{year}{2024}\natexlab{b}.
\newblock \bibinfo{title}{Mixtral of Experts}.
\newblock \bibinfo{howpublished}{\url{https://mistral.ai/news/mixtral-of-experts/}}.
\newblock


\bibitem[Bacchelli and Bird(2013)]%
        {bacchelli2013expectations}
\bibfield{author}{\bibinfo{person}{Alberto Bacchelli} {and} \bibinfo{person}{Christian Bird}.} \bibinfo{year}{2013}\natexlab{}.
\newblock \showarticletitle{Expectations, outcomes, and challenges of modern code review}. In \bibinfo{booktitle}{\emph{2013 35th International Conference on Software Engineering (ICSE)}}. IEEE, \bibinfo{pages}{712--721}.
\newblock


\bibitem[Bavota and Russo(2015)]%
        {bavota2015four}
\bibfield{author}{\bibinfo{person}{Gabriele Bavota} {and} \bibinfo{person}{Barbara Russo}.} \bibinfo{year}{2015}\natexlab{}.
\newblock \showarticletitle{Four eyes are better than two: On the impact of code reviews on software quality}. In \bibinfo{booktitle}{\emph{2015 IEEE International Conference on Software Maintenance and Evolution (ICSME)}}. IEEE, \bibinfo{pages}{81--90}.
\newblock


\bibitem[Bosu et~al\mbox{.}(2015)]%
        {bosu2015characteristics}
\bibfield{author}{\bibinfo{person}{Amiangshu Bosu}, \bibinfo{person}{Michaela Greiler}, {and} \bibinfo{person}{Christian Bird}.} \bibinfo{year}{2015}\natexlab{}.
\newblock \showarticletitle{Characteristics of useful code reviews: An empirical study at microsoft}. In \bibinfo{booktitle}{\emph{2015 IEEE/ACM 12th Working Conference on Mining Software Repositories}}. IEEE, \bibinfo{pages}{146--156}.
\newblock


\bibitem[Bouraffa and Maalej(2020)]%
        {bouraffa2020two}
\bibfield{author}{\bibinfo{person}{Abir Bouraffa} {and} \bibinfo{person}{Walid Maalej}.} \bibinfo{year}{2020}\natexlab{}.
\newblock \showarticletitle{Two decades of empirical research on developers' information needs: A preliminary analysis}. In \bibinfo{booktitle}{\emph{Proceedings of the IEEE/ACM 42nd International Conference on Software Engineering Workshops}}. \bibinfo{pages}{71--77}.
\newblock


\bibitem[Chen et~al\mbox{.}(2020)]%
        {chen2020stay}
\bibfield{author}{\bibinfo{person}{Songqiang Chen}, \bibinfo{person}{Xiaoyuan Xie}, \bibinfo{person}{Bangguo Yin}, \bibinfo{person}{Yuanxiang Ji}, \bibinfo{person}{Lin Chen}, {and} \bibinfo{person}{Baowen Xu}.} \bibinfo{year}{2020}\natexlab{}.
\newblock \showarticletitle{Stay professional and efficient: automatically generate titles for your bug reports}. In \bibinfo{booktitle}{\emph{Proceedings of the 35th IEEE/ACM International Conference on Automated Software Engineering}}. \bibinfo{pages}{385--397}.
\newblock


\bibitem[Codeflow(2016)]%
        {codeflow}
\bibfield{author}{\bibinfo{person}{Codeflow}.} \bibinfo{year}{2016}\natexlab{}.
\newblock \bibinfo{title}{Codeflow | Modern, Light-weight API integration platform.}
\newblock \bibinfo{howpublished}{\url{https://codeflow.co/}}.
\newblock
\newblock
\shownote{(Accessed on 08/02/2023)}.


\bibitem[David and Shapiro(2008)]%
        {David2008}
\bibfield{author}{\bibinfo{person}{Paul~A. David} {and} \bibinfo{person}{Joseph~S. Shapiro}.} \bibinfo{year}{2008}\natexlab{}.
\newblock \showarticletitle{Community-based production of open-source software: What do we know about the developers who participate?}
\newblock \bibinfo{journal}{\emph{Information Economics and Policy}} \bibinfo{volume}{20}, \bibinfo{number}{4} (\bibinfo{year}{2008}), \bibinfo{pages}{364--398}.
\newblock
\urldef\tempurl%
\url{https://doi.org/10.1016/j.infoecopol.2008.10.001}
\showDOI{\tempurl}


\bibitem[Ebert et~al\mbox{.}(2017)]%
        {ebert2017confusion}
\bibfield{author}{\bibinfo{person}{Felipe Ebert}, \bibinfo{person}{Fernando Castor}, \bibinfo{person}{Nicole Novielli}, {and} \bibinfo{person}{Alexander Serebrenik}.} \bibinfo{year}{2017}\natexlab{}.
\newblock \showarticletitle{Confusion detection in code reviews}. In \bibinfo{booktitle}{\emph{2017 IEEE International Conference on Software Maintenance and Evolution (ICSME)}}. IEEE, \bibinfo{pages}{549--553}.
\newblock


\bibitem[Gattrell et~al\mbox{.}(2024)]%
        {gattrell2024accord}
\bibfield{author}{\bibinfo{person}{William~T Gattrell}, \bibinfo{person}{Patricia Logullo}, \bibinfo{person}{Esther~J van Zuuren}, \bibinfo{person}{Amy Price}, \bibinfo{person}{Ellen~L Hughes}, \bibinfo{person}{Paul Blazey}, \bibinfo{person}{Christopher~C Winchester}, \bibinfo{person}{David Tovey}, \bibinfo{person}{Keith Goldman}, \bibinfo{person}{Amrit~Pali Hungin}, {et~al\mbox{.}}} \bibinfo{year}{2024}\natexlab{}.
\newblock \showarticletitle{ACCORD (ACcurate COnsensus Reporting Document): A reporting guideline for consensus methods in biomedicine developed via a modified Delphi}.
\newblock \bibinfo{journal}{\emph{PLoS medicine}} \bibinfo{volume}{21}, \bibinfo{number}{1} (\bibinfo{year}{2024}), \bibinfo{pages}{e1004326}.
\newblock


\bibitem[Geng et~al\mbox{.}(2024)]%
        {geng2024large}
\bibfield{author}{\bibinfo{person}{M. Geng}, \bibinfo{person}{S. Wang}, \bibinfo{person}{D. Dong}, \bibinfo{person}{H. Wang}, \bibinfo{person}{G. Li}, \bibinfo{person}{Z. Jin}, \bibinfo{person}{X. Mao}, {and} \bibinfo{person}{X. Liao}.} \bibinfo{year}{2024}\natexlab{}.
\newblock \showarticletitle{Large language models are few-shot summarizers: Multi-intent comment generation via in-context learning}. In \bibinfo{booktitle}{\emph{Proceedings of the 46th IEEE/ACM International Conference on Software Engineering}}. \bibinfo{pages}{1--13}.
\newblock


\bibitem[Github(2023)]%
        {github}
\bibfield{author}{\bibinfo{person}{Github}.} \bibinfo{year}{2023}\natexlab{}.
\newblock \bibinfo{title}{About pull request reviews - GitHub Docs}.
\newblock
\newblock
\urldef\tempurl%
\url{https://docs.github.com/en/pull-requests/collaborating-with-pull-requests/reviewing-changes-in-pull-requests/about-pull-request-reviews}
\showURL{%
\tempurl}
\newblock
\shownote{(Accessed on 08/02/2023)}.


\bibitem[Google(2019)]%
        {google_guide}
\bibfield{author}{\bibinfo{person}{Google}.} \bibinfo{year}{2019}\natexlab{}.
\newblock \bibinfo{title}{How to write code review comments | eng-practices}.
\newblock \bibinfo{howpublished}{\url{https://google.github.io/eng-practices/review/reviewer/comments.html}}.
\newblock
\newblock
\shownote{(Accessed on 08/02/2023)}.


\bibitem[Google(2023a)]%
        {android_cr}
\bibfield{author}{\bibinfo{person}{Google}.} \bibinfo{year}{2023}\natexlab{a}.
\newblock \bibinfo{title}{Android Gerrit}.
\newblock \bibinfo{howpublished}{\url{https://android-review.googlesource.com}}.
\newblock
\newblock
\shownote{(Accessed on 08/02/2023)}.


\bibitem[Google(2023b)]%
        {bazel_cr}
\bibfield{author}{\bibinfo{person}{Google}.} \bibinfo{year}{2023}\natexlab{b}.
\newblock \bibinfo{title}{Bazel Gerrit}.
\newblock \bibinfo{howpublished}{\url{https://bazel-review.googlesource.com}}.
\newblock
\newblock
\shownote{(Accessed on 08/02/2023)}.


\bibitem[Google(2023c)]%
        {chromium_cr}
\bibfield{author}{\bibinfo{person}{Google}.} \bibinfo{year}{2023}\natexlab{c}.
\newblock \bibinfo{title}{Chromium Gerrit}.
\newblock \bibinfo{howpublished}{\url{https://chromium-review.googlesource.com}}.
\newblock
\newblock
\shownote{(Accessed on 08/02/2023)}.


\bibitem[Google(2023d)]%
        {dart_cr}
\bibfield{author}{\bibinfo{person}{Google}.} \bibinfo{year}{2023}\natexlab{d}.
\newblock \bibinfo{title}{Dart Gerrit}.
\newblock \bibinfo{howpublished}{\url{https://dart-review.googlesource.com}}.
\newblock
\newblock
\shownote{(Accessed on 08/02/2023)}.


\bibitem[Google(2023e)]%
        {flutter_cr}
\bibfield{author}{\bibinfo{person}{Google}.} \bibinfo{year}{2023}\natexlab{e}.
\newblock \bibinfo{title}{Flutter Gerrit}.
\newblock \bibinfo{howpublished}{\url{https://flutter-review.googlesource.com}}.
\newblock
\newblock
\shownote{(Accessed on 08/02/2023)}.


\bibitem[Google(2023f)]%
        {fuschia_cr}
\bibfield{author}{\bibinfo{person}{Google}.} \bibinfo{year}{2023}\natexlab{f}.
\newblock \bibinfo{title}{Fuschia Gerrit}.
\newblock \bibinfo{howpublished}{\url{https://fuschia-review.googlesource.com}}.
\newblock
\newblock
\shownote{(Accessed on 08/02/2023)}.


\bibitem[Google(2023g)]%
        {gerrit_cr}
\bibfield{author}{\bibinfo{person}{Google}.} \bibinfo{year}{2023}\natexlab{g}.
\newblock \bibinfo{title}{Gerrit}.
\newblock \bibinfo{howpublished}{\url{https://gerrit-review.googlesource.com}}.
\newblock
\newblock
\shownote{(Accessed on 08/02/2023)}.


\bibitem[{Google}(2023)]%
        {gerrit}
\bibfield{author}{\bibinfo{person}{{Google}}.} \bibinfo{year}{2023}\natexlab{}.
\newblock \bibinfo{title}{Gerrit: Google Open Source Projects}.
\newblock \bibinfo{howpublished}{\url{https://opensource.google/projects/gerrit}}.
\newblock
\newblock
\shownote{(Accessed on 08/02/2023)}.


\bibitem[Google(2023)]%
        {go_cr}
\bibfield{author}{\bibinfo{person}{Google}.} \bibinfo{year}{2023}\natexlab{}.
\newblock \bibinfo{title}{Go Gerrit}.
\newblock \bibinfo{howpublished}{\url{https://go-review.googlesource.com}}.
\newblock
\newblock
\shownote{(Accessed on 08/02/2023)}.


\bibitem[Gunawardena et~al\mbox{.}(2023)]%
        {gunawardena2023concerns}
\bibfield{author}{\bibinfo{person}{Sanuri Gunawardena}, \bibinfo{person}{Ewan Tempero}, {and} \bibinfo{person}{Kelly Blincoe}.} \bibinfo{year}{2023}\natexlab{}.
\newblock \showarticletitle{Concerns identified in code review: A fine-grained, faceted classification}.
\newblock \bibinfo{journal}{\emph{Information and Software Technology}}  \bibinfo{volume}{153} (\bibinfo{year}{2023}), \bibinfo{pages}{107054}.
\newblock


\bibitem[Guo et~al\mbox{.}(2024)]%
        {guo2024exploring}
\bibfield{author}{\bibinfo{person}{Qi Guo}, \bibinfo{person}{Junming Cao}, \bibinfo{person}{Xiaofei Xie}, \bibinfo{person}{Shangqing Liu}, \bibinfo{person}{Xiaohong Li}, \bibinfo{person}{Bihuan Chen}, {and} \bibinfo{person}{Xin Peng}.} \bibinfo{year}{2024}\natexlab{}.
\newblock \showarticletitle{Exploring the potential of chatgpt in automated code refinement: An empirical study}. In \bibinfo{booktitle}{\emph{Proceedings of the 46th IEEE/ACM International Conference on Software Engineering}}. \bibinfo{pages}{1--13}.
\newblock


\bibitem[Gupta and Sundaresan(2018)]%
        {gupta2018intelligent}
\bibfield{author}{\bibinfo{person}{Anshul Gupta} {and} \bibinfo{person}{Neel Sundaresan}.} \bibinfo{year}{2018}\natexlab{}.
\newblock \showarticletitle{Intelligent code reviews using deep learning}. In \bibinfo{booktitle}{\emph{Proceedings of the 24th ACM SIGKDD International Conference on Knowledge Discovery and Data Mining (KDD’18) Deep Learning Day}}.
\newblock


\bibitem[Happel and Maalej(2008)]%
        {happel2008potentials}
\bibfield{author}{\bibinfo{person}{Hans-J{\"o}rg Happel} {and} \bibinfo{person}{Walid Maalej}.} \bibinfo{year}{2008}\natexlab{}.
\newblock \showarticletitle{Potentials and challenges of recommendation systems for software development}. In \bibinfo{booktitle}{\emph{Proceedings of the 2008 international workshop on Recommendation systems for software engineering}}. \bibinfo{pages}{11--15}.
\newblock


\bibitem[Hasan et~al\mbox{.}(2021)]%
        {hasan2021using}
\bibfield{author}{\bibinfo{person}{Masum Hasan}, \bibinfo{person}{Anindya Iqbal}, \bibinfo{person}{Mohammad Rafid~Ul Islam}, \bibinfo{person}{AJM~Imtiajur Rahman}, {and} \bibinfo{person}{Amiangshu Bosu}.} \bibinfo{year}{2021}\natexlab{}.
\newblock \showarticletitle{Using a balanced scorecard to identify opportunities to improve code review effectiveness: An industrial experience report}.
\newblock \bibinfo{journal}{\emph{Empirical Software Engineering}}  \bibinfo{volume}{26} (\bibinfo{year}{2021}), \bibinfo{pages}{1--34}.
\newblock


\bibitem[Hong et~al\mbox{.}(2022)]%
        {hong2022commentfinder}
\bibfield{author}{\bibinfo{person}{Yang Hong}, \bibinfo{person}{Chakkrit Tantithamthavorn}, \bibinfo{person}{Patanamon Thongtanunam}, {and} \bibinfo{person}{Aldeida Aleti}.} \bibinfo{year}{2022}\natexlab{}.
\newblock \showarticletitle{Commentfinder: a simpler, faster, more accurate code review comments recommendation}. In \bibinfo{booktitle}{\emph{Proceedings of the 30th ACM joint European software engineering conference and symposium on the foundations of software engineering}}. \bibinfo{pages}{507--519}.
\newblock


\bibitem[Huang et~al\mbox{.}(2023)]%
        {huang2023ischatgpt}
\bibfield{author}{\bibinfo{person}{Fan Huang}, \bibinfo{person}{Haewoon Kwak}, {and} \bibinfo{person}{Jisun An}.} \bibinfo{year}{2023}\natexlab{}.
\newblock \showarticletitle{Is ChatGPT Better than Human Annotators? Potential and Limitations of ChatGPT in Explaining Implicit Hate Speech}. In \bibinfo{booktitle}{\emph{Companion Proceedings of the ACM Web Conference 2023}} (Austin, TX, USA) \emph{(\bibinfo{series}{WWW '23 Companion})}. \bibinfo{publisher}{Association for Computing Machinery}, \bibinfo{address}{New York, NY, USA}, \bibinfo{pages}{294–297}.
\newblock
\showISBNx{9781450394192}
\urldef\tempurl%
\url{https://doi.org/10.1145/3543873.3587368}
\showDOI{\tempurl}


\bibitem[Huq et~al\mbox{.}(2022)]%
        {huq2022review4repair}
\bibfield{author}{\bibinfo{person}{Faria Huq}, \bibinfo{person}{Masum Hasan}, \bibinfo{person}{Md~Mahim~Anjum Haque}, \bibinfo{person}{Sazan Mahbub}, \bibinfo{person}{Anindya Iqbal}, {and} \bibinfo{person}{Toufique Ahmed}.} \bibinfo{year}{2022}\natexlab{}.
\newblock \showarticletitle{Review4Repair: Code review aided automatic program repairing}.
\newblock \bibinfo{journal}{\emph{Information and Software Technology}}  \bibinfo{volume}{143} (\bibinfo{year}{2022}), \bibinfo{pages}{106765}.
\newblock


\bibitem[Jiang et~al\mbox{.}(2024)]%
        {jiang2024evaluating}
\bibfield{author}{\bibinfo{person}{Shengbei Jiang}, \bibinfo{person}{Jiabao Zhang}, \bibinfo{person}{Wei Chen}, \bibinfo{person}{Bo Wang}, \bibinfo{person}{Jianyi Zhou}, {and} \bibinfo{person}{Jie~M Zhang}.} \bibinfo{year}{2024}\natexlab{}.
\newblock \showarticletitle{Evaluating Fault Localization and Program Repair Capabilities of Existing Closed-Source General-Purpose LLMs}.
\newblock  (\bibinfo{year}{2024}).
\newblock


\bibitem[Kamienski and Bezemer(2021)]%
        {Kamienski2021}
\bibfield{author}{\bibinfo{person}{Arthur Kamienski} {and} \bibinfo{person}{Cor-Paul Bezemer}.} \bibinfo{year}{2021}\natexlab{}.
\newblock \showarticletitle{An empirical study of Q\&A websites for game developers}.
\newblock \bibinfo{journal}{\emph{Empirical Software Engineering}} \bibinfo{volume}{26}, \bibinfo{number}{6} (\bibinfo{year}{2021}), \bibinfo{pages}{115}.
\newblock
\urldef\tempurl%
\url{https://doi.org/10.1007/s10664-021-10014-4}
\showDOI{\tempurl}


\bibitem[Kang et~al\mbox{.}(2024)]%
        {kang2024human}
\bibfield{author}{\bibinfo{person}{Hong~Jin Kang}, \bibinfo{person}{Fabrice Harel-Canada}, \bibinfo{person}{Muhammad~Ali Gulzar}, \bibinfo{person}{Violet Peng}, {and} \bibinfo{person}{Miryung Kim}.} \bibinfo{year}{2024}\natexlab{}.
\newblock \showarticletitle{Human-in-the-Loop Synthetic Text Data Inspection with Provenance Tracking}. In \bibinfo{booktitle}{\emph{NAACL 2024}}.
\newblock


\bibitem[Khatoonabadi et~al\mbox{.}(2023)]%
        {khatoonabadi2023wasted}
\bibfield{author}{\bibinfo{person}{SayedHassan Khatoonabadi}, \bibinfo{person}{Diego~Elias Costa}, \bibinfo{person}{Rabe Abdalkareem}, {and} \bibinfo{person}{Emad Shihab}.} \bibinfo{year}{2023}\natexlab{}.
\newblock \showarticletitle{On wasted contributions: understanding the dynamics of contributor-abandoned pull requests--A mixed-methods study of 10 large open-source projects}.
\newblock \bibinfo{journal}{\emph{ACM Transactions on Software Engineering and Methodology}} \bibinfo{volume}{32}, \bibinfo{number}{1} (\bibinfo{year}{2023}), \bibinfo{pages}{1--39}.
\newblock


\bibitem[Kononenko et~al\mbox{.}(2016)]%
        {kononenko2016code}
\bibfield{author}{\bibinfo{person}{Oleksii Kononenko}, \bibinfo{person}{Olga Baysal}, {and} \bibinfo{person}{Michael~W Godfrey}.} \bibinfo{year}{2016}\natexlab{}.
\newblock \showarticletitle{Code review quality: How developers see it}. In \bibinfo{booktitle}{\emph{Proceedings of the 38th international conference on software engineering}}. \bibinfo{pages}{1028--1038}.
\newblock


\bibitem[Kononenko et~al\mbox{.}(2015)]%
        {kononenko2015investigating}
\bibfield{author}{\bibinfo{person}{Oleksii Kononenko}, \bibinfo{person}{Olga Baysal}, \bibinfo{person}{Latifa Guerrouj}, \bibinfo{person}{Yaxin Cao}, {and} \bibinfo{person}{Michael~W Godfrey}.} \bibinfo{year}{2015}\natexlab{}.
\newblock \showarticletitle{Investigating code review quality: Do people and participation matter?}. In \bibinfo{booktitle}{\emph{2015 IEEE international conference on software maintenance and evolution (ICSME)}}. IEEE, \bibinfo{pages}{111--120}.
\newblock


\bibitem[Kosti et~al\mbox{.}(2014)]%
        {Kosti2014}
\bibfield{author}{\bibinfo{person}{Makrina~Viola Kosti}, \bibinfo{person}{Robert Feldt}, {and} \bibinfo{person}{Lefteris Angelis}.} \bibinfo{year}{2014}\natexlab{}.
\newblock \showarticletitle{Personality, emotional intelligence and work preferences in software engineering: An empirical study}.
\newblock \bibinfo{journal}{\emph{Information and Software Technology}} \bibinfo{volume}{56}, \bibinfo{number}{8} (\bibinfo{year}{2014}), \bibinfo{pages}{973--990}.
\newblock
\urldef\tempurl%
\url{https://doi.org/10.1016/j.infsof.2014.03.004}
\showDOI{\tempurl}


\bibitem[Li et~al\mbox{.}(2022b)]%
        {li2022auger}
\bibfield{author}{\bibinfo{person}{Lingwei Li}, \bibinfo{person}{Li Yang}, \bibinfo{person}{Huaxi Jiang}, \bibinfo{person}{Jun Yan}, \bibinfo{person}{Tiejian Luo}, \bibinfo{person}{Zihan Hua}, \bibinfo{person}{Geng Liang}, {and} \bibinfo{person}{Chun Zuo}.} \bibinfo{year}{2022}\natexlab{b}.
\newblock \showarticletitle{AUGER: automatically generating review comments with pre-training models}. In \bibinfo{booktitle}{\emph{Proceedings of the 30th ACM Joint European Software Engineering Conference and Symposium on the Foundations of Software Engineering}}. \bibinfo{pages}{1009--1021}.
\newblock


\bibitem[Li et~al\mbox{.}(2022a)]%
        {li2022automating}
\bibfield{author}{\bibinfo{person}{Zhiyu Li}, \bibinfo{person}{Shuai Lu}, \bibinfo{person}{Daya Guo}, \bibinfo{person}{Nan Duan}, \bibinfo{person}{Shailesh Jannu}, \bibinfo{person}{Grant Jenks}, \bibinfo{person}{Deep Majumder}, \bibinfo{person}{Jared Green}, \bibinfo{person}{Alexey Svyatkovskiy}, \bibinfo{person}{Shengyu Fu}, {et~al\mbox{.}}} \bibinfo{year}{2022}\natexlab{a}.
\newblock \showarticletitle{Automating code review activities by large-scale pre-training}. In \bibinfo{booktitle}{\emph{Proceedings of the 30th ACM Joint European Software Engineering Conference and Symposium on the Foundations of Software Engineering}}. \bibinfo{pages}{1035--1047}.
\newblock


\bibitem[Liang et~al\mbox{.}(2019)]%
        {liang2019explain}
\bibfield{author}{\bibinfo{person}{Jingjing Liang}, \bibinfo{person}{Yaozong Hou}, \bibinfo{person}{Shurui Zhou}, \bibinfo{person}{Junjie Chen}, \bibinfo{person}{Yingfei Xiong}, {and} \bibinfo{person}{Gang Huang}.} \bibinfo{year}{2019}\natexlab{}.
\newblock \showarticletitle{How to explain a patch: An empirical study of patch explanations in open source projects}. In \bibinfo{booktitle}{\emph{2019 IEEE 30th International Symposium on Software Reliability Engineering (ISSRE)}}. IEEE, \bibinfo{pages}{58--69}.
\newblock


\bibitem[Lim et~al\mbox{.}(2019)]%
        {lim2019these}
\bibfield{author}{\bibinfo{person}{Brian~Y Lim}, \bibinfo{person}{Qian Yang}, \bibinfo{person}{Ashraf~M Abdul}, {and} \bibinfo{person}{Danding Wang}.} \bibinfo{year}{2019}\natexlab{}.
\newblock \showarticletitle{Why these explanations? Selecting intelligibility types for explanation goals.}. In \bibinfo{booktitle}{\emph{IUI Workshops}}.
\newblock


\bibitem[Liu et~al\mbox{.}(2024)]%
        {liu2024your}
\bibfield{author}{\bibinfo{person}{Jiawei Liu}, \bibinfo{person}{Chunqiu~Steven Xia}, \bibinfo{person}{Yuyao Wang}, {and} \bibinfo{person}{Lingming Zhang}.} \bibinfo{year}{2024}\natexlab{}.
\newblock \showarticletitle{Is your code generated by chatgpt really correct? rigorous evaluation of large language models for code generation}.
\newblock \bibinfo{journal}{\emph{Advances in Neural Information Processing Systems}}  \bibinfo{volume}{36} (\bibinfo{year}{2024}).
\newblock


\bibitem[Liu et~al\mbox{.}(2023)]%
        {liu2023refining}
\bibfield{author}{\bibinfo{person}{Yue Liu}, \bibinfo{person}{Thanh Le-Cong}, \bibinfo{person}{Ratnadira Widyasari}, \bibinfo{person}{Chakkrit Tantithamthavorn}, \bibinfo{person}{Li Li}, \bibinfo{person}{Xuan-Bach~D Le}, {and} \bibinfo{person}{David Lo}.} \bibinfo{year}{2023}\natexlab{}.
\newblock \showarticletitle{Refining ChatGPT-generated code: Characterizing and mitigating code quality issues}.
\newblock \bibinfo{journal}{\emph{ACM Transactions on Software Engineering and Methodology}} (\bibinfo{year}{2023}).
\newblock


\bibitem[Maalej(2023)]%
        {Maalej:BotSE:2023}
\bibfield{author}{\bibinfo{person}{Walid Maalej}.} \bibinfo{year}{2023}\natexlab{}.
\newblock \showarticletitle{From RSSE to BotSE: Potentials and Challenges Revisited after 15 Years}. In \bibinfo{booktitle}{\emph{2023 IEEE/ACM 5th International Workshop on Bots in Software Engineering (BotSE)}}. \bibinfo{pages}{19--22}.
\newblock
\urldef\tempurl%
\url{https://doi.org/10.1109/BotSE59190.2023.00012}
\showDOI{\tempurl}


\bibitem[Maalej and Robillard(2013)]%
        {maalej2013patterns}
\bibfield{author}{\bibinfo{person}{Walid Maalej} {and} \bibinfo{person}{Martin~P Robillard}.} \bibinfo{year}{2013}\natexlab{}.
\newblock \showarticletitle{Patterns of knowledge in API reference documentation}.
\newblock \bibinfo{journal}{\emph{IEEE Transactions on software Engineering}} \bibinfo{volume}{39}, \bibinfo{number}{9} (\bibinfo{year}{2013}), \bibinfo{pages}{1264--1282}.
\newblock


\bibitem[Maalej et~al\mbox{.}(2014)]%
        {maalej2014comprehension}
\bibfield{author}{\bibinfo{person}{Walid Maalej}, \bibinfo{person}{Rebecca Tiarks}, \bibinfo{person}{Tobias Roehm}, {and} \bibinfo{person}{Rainer Koschke}.} \bibinfo{year}{2014}\natexlab{}.
\newblock \showarticletitle{On the comprehension of program comprehension}.
\newblock \bibinfo{journal}{\emph{ACM Transactions on Software Engineering and Methodology (TOSEM)}} \bibinfo{volume}{23}, \bibinfo{number}{4} (\bibinfo{year}{2014}), \bibinfo{pages}{1--37}.
\newblock


\bibitem[MacNeil et~al\mbox{.}(2023)]%
        {macneil2023experiences}
\bibfield{author}{\bibinfo{person}{Stephen MacNeil}, \bibinfo{person}{Andrew Tran}, \bibinfo{person}{Arto Hellas}, \bibinfo{person}{Joanne Kim}, \bibinfo{person}{Sami Sarsa}, \bibinfo{person}{Paul Denny}, \bibinfo{person}{Seth Bernstein}, {and} \bibinfo{person}{Juho Leinonen}.} \bibinfo{year}{2023}\natexlab{}.
\newblock \showarticletitle{Experiences from using code explanations generated by large language models in a web software development e-book}. In \bibinfo{booktitle}{\emph{Proceedings of the 54th ACM Technical Symposium on Computer Science Education V. 1}}. \bibinfo{pages}{931--937}.
\newblock


\bibitem[McIntosh et~al\mbox{.}(2014)]%
        {mcintosh2014impact}
\bibfield{author}{\bibinfo{person}{Shane McIntosh}, \bibinfo{person}{Yasutaka Kamei}, \bibinfo{person}{Bram Adams}, {and} \bibinfo{person}{Ahmed~E Hassan}.} \bibinfo{year}{2014}\natexlab{}.
\newblock \showarticletitle{The impact of code review coverage and code review participation on software quality: A case study of the qt, vtk, and itk projects}. In \bibinfo{booktitle}{\emph{Proceedings of the 11th working conference on mining software repositories}}. \bibinfo{pages}{192--201}.
\newblock


\bibitem[Morales et~al\mbox{.}(2015)]%
        {morales2015code}
\bibfield{author}{\bibinfo{person}{Rodrigo Morales}, \bibinfo{person}{Shane McIntosh}, {and} \bibinfo{person}{Foutse Khomh}.} \bibinfo{year}{2015}\natexlab{}.
\newblock \showarticletitle{Do code review practices impact design quality? a case study of the qt, vtk, and itk projects}. In \bibinfo{booktitle}{\emph{2015 IEEE 22nd international conference on software analysis, evolution, and reengineering (SANER)}}. IEEE, \bibinfo{pages}{171--180}.
\newblock


\bibitem[OpenAI(2023)]%
        {gptversion}
\bibfield{author}{\bibinfo{person}{OpenAI}.} \bibinfo{year}{2023}\natexlab{}.
\newblock \bibinfo{title}{ChatGPT Version}.
\newblock \bibinfo{howpublished}{\url{https://help.openai.com/en/articles/6825453-chatgpt-release-notes}}.
\newblock
\newblock
\shownote{(Accessed on 08/02/2023)}.


\bibitem[Paixao et~al\mbox{.}(2018)]%
        {paixao2018crop}
\bibfield{author}{\bibinfo{person}{Matheus Paixao}, \bibinfo{person}{Jens Krinke}, \bibinfo{person}{Donggyun Han}, {and} \bibinfo{person}{Mark Harman}.} \bibinfo{year}{2018}\natexlab{}.
\newblock \showarticletitle{CROP: Linking code reviews to source code changes}. In \bibinfo{booktitle}{\emph{Proceedings of the 15th international conference on mining software repositories}}. \bibinfo{pages}{46--49}.
\newblock


\bibitem[Pascarella et~al\mbox{.}(2018)]%
        {pascarella2018information}
\bibfield{author}{\bibinfo{person}{Luca Pascarella}, \bibinfo{person}{Davide Spadini}, \bibinfo{person}{Fabio Palomba}, \bibinfo{person}{Magiel Bruntink}, {and} \bibinfo{person}{Alberto Bacchelli}.} \bibinfo{year}{2018}\natexlab{}.
\newblock \showarticletitle{Information needs in contemporary code review}.
\newblock \bibinfo{journal}{\emph{Proceedings of the ACM on Human-Computer Interaction}} \bibinfo{volume}{2}, \bibinfo{number}{CSCW} (\bibinfo{year}{2018}), \bibinfo{pages}{1--27}.
\newblock


\bibitem[Phacility(2021)]%
        {phacility}
\bibfield{author}{\bibinfo{person}{Phacility}.} \bibinfo{year}{2021}\natexlab{}.
\newblock \bibinfo{title}{Phacility - Phabricator}.
\newblock
\newblock
\urldef\tempurl%
\url{https://phacility.com/phabricator/}
\showURL{%
\tempurl}
\newblock
\shownote{(Accessed on 08/02/2023)}.


\bibitem[Quirk et~al\mbox{.}(2012)]%
        {quirk2012msr}
\bibfield{author}{\bibinfo{person}{Chris Quirk}, \bibinfo{person}{Pallavi Choudhury}, \bibinfo{person}{Jianfeng Gao}, \bibinfo{person}{Hisami Suzuki}, \bibinfo{person}{Kristina Toutanova}, \bibinfo{person}{Michael Gamon}, \bibinfo{person}{Scott Wen-tau Yih}, \bibinfo{person}{Lucy Vanderwende}, {and} \bibinfo{person}{Colin Cherry}.} \bibinfo{year}{2012}\natexlab{}.
\newblock \showarticletitle{MSR SPLAT, a language analysis toolkit}. In \bibinfo{booktitle}{\emph{Proceedings of NAACL-HLT 2012}}.
\newblock


\bibitem[Rahman et~al\mbox{.}(2017)]%
        {rahman2017predicting}
\bibfield{author}{\bibinfo{person}{Mohammad~Masudur Rahman}, \bibinfo{person}{Chanchal~K Roy}, {and} \bibinfo{person}{Raula~G Kula}.} \bibinfo{year}{2017}\natexlab{}.
\newblock \showarticletitle{Predicting usefulness of code review comments using textual features and developer experience}. In \bibinfo{booktitle}{\emph{2017 IEEE/ACM 14th International Conference on Mining Software Repositories (MSR)}}. IEEE, \bibinfo{pages}{215--226}.
\newblock


\bibitem[Rahman et~al\mbox{.}(2022)]%
        {rahman2022example}
\bibfield{author}{\bibinfo{person}{Shadikur Rahman}, \bibinfo{person}{Umme~Ayman Koana}, {and} \bibinfo{person}{Maleknaz Nayebi}.} \bibinfo{year}{2022}\natexlab{}.
\newblock \showarticletitle{Example Driven Code Review Explanation}. In \bibinfo{booktitle}{\emph{Proceedings of the 16th ACM/IEEE International Symposium on Empirical Software Engineering and Measurement}}. \bibinfo{pages}{307--312}.
\newblock


\bibitem[Research(2024)]%
        {Google_Gemini}
\bibfield{author}{\bibinfo{person}{Google Research}.} \bibinfo{year}{2024}\natexlab{}.
\newblock \bibinfo{title}{Google Gemini}.
\newblock \bibinfo{howpublished}{\url{https://gemini.google.com}}.
\newblock


\bibitem[Rodr{\'\i}guez-Ma{\~n}as et~al\mbox{.}(2013)]%
        {rodriguez2013searching}
\bibfield{author}{\bibinfo{person}{Leocadio Rodr{\'\i}guez-Ma{\~n}as}, \bibinfo{person}{Catherine F{\'e}art}, \bibinfo{person}{Giovanni Mann}, \bibinfo{person}{Jose Vi{\~n}a}, \bibinfo{person}{Somnath Chatterji}, \bibinfo{person}{Wojtek Chodzko-Zajko}, \bibinfo{person}{Magali Gonzalez-Cola{\c{c}}o~Harmand}, \bibinfo{person}{Howard Bergman}, \bibinfo{person}{Laure Carcaillon}, \bibinfo{person}{Caroline Nicholson}, {et~al\mbox{.}}} \bibinfo{year}{2013}\natexlab{}.
\newblock \showarticletitle{Searching for an operational definition of frailty: a Delphi method based consensus statement. The frailty operative definition-consensus conference project}.
\newblock \bibinfo{journal}{\emph{Journals of gerontology series a: biomedical sciences and medical sciences}} \bibinfo{volume}{68}, \bibinfo{number}{1} (\bibinfo{year}{2013}), \bibinfo{pages}{62--67}.
\newblock


\bibitem[Sadowski et~al\mbox{.}(2018)]%
        {sadowski2018modern}
\bibfield{author}{\bibinfo{person}{Caitlin Sadowski}, \bibinfo{person}{Emma S{\"o}derberg}, \bibinfo{person}{Luke Church}, \bibinfo{person}{Michal Sipko}, {and} \bibinfo{person}{Alberto Bacchelli}.} \bibinfo{year}{2018}\natexlab{}.
\newblock \showarticletitle{Modern code review: a case study at google}. In \bibinfo{booktitle}{\emph{Proceedings of the 40th international conference on software engineering: Software engineering in practice}}. \bibinfo{pages}{181--190}.
\newblock


\bibitem[Spencer(2009)]%
        {spencer2009card}
\bibfield{author}{\bibinfo{person}{Donna Spencer}.} \bibinfo{year}{2009}\natexlab{}.
\newblock \bibinfo{booktitle}{\emph{Card sorting: Designing usable categories}}.
\newblock \bibinfo{publisher}{Rosenfeld Media}.
\newblock


\bibitem[Stanik et~al\mbox{.}(2018)]%
        {Stanik:ICSME:2018}
\bibfield{author}{\bibinfo{person}{Christoph Stanik}, \bibinfo{person}{Lloyd Montgomery}, \bibinfo{person}{Daniel Martens}, \bibinfo{person}{Davide Fucci}, {and} \bibinfo{person}{Walid Maalej}.} \bibinfo{year}{2018}\natexlab{}.
\newblock \showarticletitle{A simple nlp-based approach to support onboarding and retention in open source communities}. In \bibinfo{booktitle}{\emph{2018 IEEE international conference on software maintenance and evolution (ICSME)}}. IEEE, \bibinfo{pages}{172--182}.
\newblock


\bibitem[Sun et~al\mbox{.}(2024)]%
        {sun2024gptscan}
\bibfield{author}{\bibinfo{person}{Yuqiang Sun}, \bibinfo{person}{Daoyuan Wu}, \bibinfo{person}{Yue Xue}, \bibinfo{person}{Han Liu}, \bibinfo{person}{Haijun Wang}, \bibinfo{person}{Zhengzi Xu}, \bibinfo{person}{Xiaofei Xie}, {and} \bibinfo{person}{Yang Liu}.} \bibinfo{year}{2024}\natexlab{}.
\newblock \showarticletitle{Gptscan: Detecting logic vulnerabilities in smart contracts by combining gpt with program analysis}. In \bibinfo{booktitle}{\emph{Proceedings of the IEEE/ACM 46th International Conference on Software Engineering}}. \bibinfo{pages}{1--13}.
\newblock


\bibitem[Taibi et~al\mbox{.}(2017)]%
        {Taibi2017}
\bibfield{author}{\bibinfo{person}{Davide Taibi}, \bibinfo{person}{Andrea Janes}, {and} \bibinfo{person}{Valentina Lenarduzzi}.} \bibinfo{year}{2017}\natexlab{}.
\newblock \showarticletitle{How developers perceive smells in source code: A replicated study}.
\newblock \bibinfo{journal}{\emph{Information and Software Technology}}  \bibinfo{volume}{92} (\bibinfo{year}{2017}), \bibinfo{pages}{223--235}.
\newblock
\urldef\tempurl%
\url{https://doi.org/10.1016/j.infsof.2017.08.007}
\showDOI{\tempurl}


\bibitem[Tallarida et~al\mbox{.}(1987)]%
        {tallarida1987chi}
\bibfield{author}{\bibinfo{person}{Ronald~J Tallarida}, \bibinfo{person}{Rodney~B Murray}, \bibinfo{person}{Ronald~J Tallarida}, {and} \bibinfo{person}{Rodney~B Murray}.} \bibinfo{year}{1987}\natexlab{}.
\newblock \showarticletitle{Chi-square test}.
\newblock \bibinfo{journal}{\emph{Manual of pharmacologic calculations: with computer programs}} (\bibinfo{year}{1987}), \bibinfo{pages}{140--142}.
\newblock


\bibitem[Tang and Nadi(2023)]%
        {Tang:MSR:2023}
\bibfield{author}{\bibinfo{person}{Henry Tang} {and} \bibinfo{person}{Sarah Nadi}.} \bibinfo{year}{2023}\natexlab{}.
\newblock \showarticletitle{Evaluating Software Documentation Quality}. In \bibinfo{booktitle}{\emph{2023 IEEE/ACM 20th International Conference on Mining Software Repositories (MSR)}}. \bibinfo{pages}{67--78}.
\newblock
\urldef\tempurl%
\url{https://doi.org/10.1109/MSR59073.2023.00023}
\showDOI{\tempurl}


\bibitem[Thongtanunam et~al\mbox{.}(2022)]%
        {thongtanunam2022autotransform}
\bibfield{author}{\bibinfo{person}{Patanamon Thongtanunam}, \bibinfo{person}{Chanathip Pornprasit}, {and} \bibinfo{person}{Chakkrit Tantithamthavorn}.} \bibinfo{year}{2022}\natexlab{}.
\newblock \showarticletitle{AutoTransform: automated code transformation to support modern code review process}. In \bibinfo{booktitle}{\emph{Proceedings of the 44th International Conference on Software Engineering}}. \bibinfo{pages}{237--248}.
\newblock


\bibitem[Thongtanunam et~al\mbox{.}(2015)]%
        {thongtanunam2015should}
\bibfield{author}{\bibinfo{person}{Patanamon Thongtanunam}, \bibinfo{person}{Chakkrit Tantithamthavorn}, \bibinfo{person}{Raula~Gaikovina Kula}, \bibinfo{person}{Norihiro Yoshida}, \bibinfo{person}{Hajimu Iida}, {and} \bibinfo{person}{Ken-ichi Matsumoto}.} \bibinfo{year}{2015}\natexlab{}.
\newblock \showarticletitle{Who should review my code? a file location-based code-reviewer recommendation approach for modern code review}. In \bibinfo{booktitle}{\emph{2015 IEEE 22nd International Conference on Software Analysis, Evolution, and Reengineering (SANER)}}. IEEE, \bibinfo{pages}{141--150}.
\newblock


\bibitem[Tufano et~al\mbox{.}(2019)]%
        {tufano2019learning}
\bibfield{author}{\bibinfo{person}{Michele Tufano}, \bibinfo{person}{Jevgenija Pantiuchina}, \bibinfo{person}{Cody Watson}, \bibinfo{person}{Gabriele Bavota}, {and} \bibinfo{person}{Denys Poshyvanyk}.} \bibinfo{year}{2019}\natexlab{}.
\newblock \showarticletitle{On learning meaningful code changes via neural machine translation}. In \bibinfo{booktitle}{\emph{2019 IEEE/ACM 41st International Conference on Software Engineering (ICSE)}}. IEEE, \bibinfo{pages}{25--36}.
\newblock


\bibitem[Tufano et~al\mbox{.}(2022)]%
        {tufano2022using}
\bibfield{author}{\bibinfo{person}{Rosalia Tufano}, \bibinfo{person}{Simone Masiero}, \bibinfo{person}{Antonio Mastropaolo}, \bibinfo{person}{Luca Pascarella}, \bibinfo{person}{Denys Poshyvanyk}, {and} \bibinfo{person}{Gabriele Bavota}.} \bibinfo{year}{2022}\natexlab{}.
\newblock \showarticletitle{Using pre-trained models to boost code review automation}. In \bibinfo{booktitle}{\emph{Proceedings of the 44th International Conference on Software Engineering}}. \bibinfo{pages}{2291--2302}.
\newblock


\bibitem[Tufano et~al\mbox{.}(2021)]%
        {tufano2021towards}
\bibfield{author}{\bibinfo{person}{Rosalia Tufano}, \bibinfo{person}{Luca Pascarella}, \bibinfo{person}{Michele Tufano}, \bibinfo{person}{Denys Poshyvanyk}, {and} \bibinfo{person}{Gabriele Bavota}.} \bibinfo{year}{2021}\natexlab{}.
\newblock \showarticletitle{Towards automating code review activities}. In \bibinfo{booktitle}{\emph{2021 IEEE/ACM 43rd International Conference on Software Engineering (ICSE)}}. IEEE, \bibinfo{pages}{163--174}.
\newblock


\bibitem[Turzo and Bosu(2023)]%
        {turzo2023makes}
\bibfield{author}{\bibinfo{person}{Asif~Kamal Turzo} {and} \bibinfo{person}{Amiangshu Bosu}.} \bibinfo{year}{2023}\natexlab{}.
\newblock \showarticletitle{What Makes a Code Review Useful to OpenDev Developers? An Empirical Investigation}.
\newblock \bibinfo{journal}{\emph{arXiv preprint arXiv:2302.11686}} (\bibinfo{year}{2023}).
\newblock


\bibitem[Uch{\^o}a et~al\mbox{.}(2021)]%
        {uchoa2021predicting}
\bibfield{author}{\bibinfo{person}{Anderson Uch{\^o}a}, \bibinfo{person}{Caio Barbosa}, \bibinfo{person}{Daniel Coutinho}, \bibinfo{person}{Willian Oizumi}, \bibinfo{person}{Wesley~KG Assun{\c{c}}ao}, \bibinfo{person}{Silvia~Regina Vergilio}, \bibinfo{person}{Juliana~Alves Pereira}, \bibinfo{person}{Anderson Oliveira}, {and} \bibinfo{person}{Alessandro Garcia}.} \bibinfo{year}{2021}\natexlab{}.
\newblock \showarticletitle{Predicting design impactful changes in modern code review: A large-scale empirical study}. In \bibinfo{booktitle}{\emph{2021 IEEE/ACM 18th International Conference on Mining Software Repositories (MSR)}}. IEEE, \bibinfo{pages}{471--482}.
\newblock


\bibitem[Wang et~al\mbox{.}(2019)]%
        {wang2019designing}
\bibfield{author}{\bibinfo{person}{Danding Wang}, \bibinfo{person}{Qian Yang}, \bibinfo{person}{Ashraf Abdul}, {and} \bibinfo{person}{Brian~Y Lim}.} \bibinfo{year}{2019}\natexlab{}.
\newblock \showarticletitle{Designing theory-driven user-centric explainable AI}. In \bibinfo{booktitle}{\emph{Proceedings of the 2019 CHI conference on human factors in computing systems}}. \bibinfo{pages}{1--15}.
\newblock


\bibitem[Widyasari et~al\mbox{.}(2024)]%
        {widyasari2024demystifying}
\bibfield{author}{\bibinfo{person}{Ratnadira Widyasari}, \bibinfo{person}{Jia~Wei Ang}, \bibinfo{person}{Truong~Giang Nguyen}, \bibinfo{person}{Neil Sharma}, {and} \bibinfo{person}{David Lo}.} \bibinfo{year}{2024}\natexlab{}.
\newblock \showarticletitle{Demystifying Faulty Code with LLM: Step-by-Step Reasoning for Explainable Fault Localization}. In \bibinfo{booktitle}{\emph{2024 IEEE international conference on software analysis, evolution, and reengineering (SANER)}}. IEEE.
\newblock


\bibitem[Widyasari et~al\mbox{.}(2023)]%
        {replication_package}
\bibfield{author}{\bibinfo{person}{Ratnadira Widyasari}, \bibinfo{person}{Ting Zhang}, \bibinfo{person}{Abir Bouraffa}, \bibinfo{person}{Walid Maalej}, {and} \bibinfo{person}{David Lo}.} \bibinfo{year}{2023}\natexlab{}.
\newblock \bibinfo{title}{Code Review Explanation Replication Package}.
\newblock \bibinfo{howpublished}{\url{https://figshare.com/s/135201b8f87ab705448b}}.
\newblock
\newblock
\shownote{(Accessed on 08/02/2023)}.


\bibitem[Wurzel~Gon{\c{c}}alves et~al\mbox{.}(2022)]%
        {wurzel2022interpersonal}
\bibfield{author}{\bibinfo{person}{Pavl{\'\i}na Wurzel~Gon{\c{c}}alves}, \bibinfo{person}{G{\"u}l {\c{C}}alikli}, {and} \bibinfo{person}{Alberto Bacchelli}.} \bibinfo{year}{2022}\natexlab{}.
\newblock \showarticletitle{Interpersonal Conflicts During Code Review: Developers' Experience and Practices}.
\newblock \bibinfo{journal}{\emph{Proceedings of the ACM on Human-Computer Interaction}} \bibinfo{volume}{6}, \bibinfo{number}{CSCW1} (\bibinfo{year}{2022}), \bibinfo{pages}{1--33}.
\newblock


\bibitem[Wurzel~Gon{\c{c}}alves et~al\mbox{.}(2023)]%
        {wurzel2023competencies}
\bibfield{author}{\bibinfo{person}{Pavl{\'\i}na Wurzel~Gon{\c{c}}alves}, \bibinfo{person}{G{\"u}l Calikli}, \bibinfo{person}{Alexander Serebrenik}, {and} \bibinfo{person}{Alberto Bacchelli}.} \bibinfo{year}{2023}\natexlab{}.
\newblock \showarticletitle{Competencies for Code Review}.
\newblock \bibinfo{journal}{\emph{Proceedings of the ACM on Human-Computer Interaction}} \bibinfo{volume}{7}, \bibinfo{number}{CSCW1} (\bibinfo{year}{2023}), \bibinfo{pages}{1--33}.
\newblock


\bibitem[Xia and Zhang(2023)]%
        {xia2023keep}
\bibfield{author}{\bibinfo{person}{Chunqiu~Steven Xia} {and} \bibinfo{person}{Lingming Zhang}.} \bibinfo{year}{2023}\natexlab{}.
\newblock \showarticletitle{Keep the Conversation Going: Fixing 162 out of 337 bugs for \$0.42 each using ChatGPT}.
\newblock \bibinfo{journal}{\emph{arXiv preprint arXiv:2304.00385}} (\bibinfo{year}{2023}).
\newblock


\bibitem[Xu et~al\mbox{.}(2020)]%
        {xu2020reinventing}
\bibfield{author}{\bibinfo{person}{Bowen Xu}, \bibinfo{person}{Le An}, \bibinfo{person}{Ferdian Thung}, \bibinfo{person}{Foutse Khomh}, {and} \bibinfo{person}{David Lo}.} \bibinfo{year}{2020}\natexlab{}.
\newblock \showarticletitle{Why reinventing the wheels? An empirical study on library reuse and re-implementation}.
\newblock \bibinfo{journal}{\emph{Empirical Software Engineering}}  \bibinfo{volume}{25} (\bibinfo{year}{2020}), \bibinfo{pages}{755--789}.
\newblock


\bibitem[Zanjani et~al\mbox{.}(2015)]%
        {zanjani2015automatically}
\bibfield{author}{\bibinfo{person}{Motahareh~Bahrami Zanjani}, \bibinfo{person}{Huzefa Kagdi}, {and} \bibinfo{person}{Christian Bird}.} \bibinfo{year}{2015}\natexlab{}.
\newblock \showarticletitle{Automatically recommending peer reviewers in modern code review}.
\newblock \bibinfo{journal}{\emph{IEEE Transactions on Software Engineering}} \bibinfo{volume}{42}, \bibinfo{number}{6} (\bibinfo{year}{2015}), \bibinfo{pages}{530--543}.
\newblock


\bibitem[Zhang et~al\mbox{.}(2020a)]%
        {soarsmu}
\bibfield{author}{\bibinfo{person}{Ting Zhang}, \bibinfo{person}{Bowen Xu}, \bibinfo{person}{Ferdian Thung}, \bibinfo{person}{Stefanus~Agus Haryono}, \bibinfo{person}{David Lo}, {and} \bibinfo{person}{Lingxiao Jiang}.} \bibinfo{year}{2020}\natexlab{a}.
\newblock \bibinfo{title}{The replication package of Sentiment analysis for software engineering: How far can pre-trained transformer models go? accepted by ICSME'20.}
\newblock \bibinfo{howpublished}{\url{https://github.com/soarsmu/SA4SE/}}.
\newblock
\newblock
\shownote{(Accessed on 08/02/2023)}.


\bibitem[Zhang et~al\mbox{.}(2020b)]%
        {zhang2020sentiment}
\bibfield{author}{\bibinfo{person}{Ting Zhang}, \bibinfo{person}{Bowen Xu}, \bibinfo{person}{Ferdian Thung}, \bibinfo{person}{Stefanus~Agus Haryono}, \bibinfo{person}{David Lo}, {and} \bibinfo{person}{Lingxiao Jiang}.} \bibinfo{year}{2020}\natexlab{b}.
\newblock \showarticletitle{Sentiment analysis for software engineering: How far can pre-trained transformer models go?}. In \bibinfo{booktitle}{\emph{2020 IEEE International Conference on Software Maintenance and Evolution (ICSME)}}. IEEE, \bibinfo{pages}{70--80}.
\newblock


\bibitem[Zhou et~al\mbox{.}(2024)]%
        {zhou2024large}
\bibfield{author}{\bibinfo{person}{Xin Zhou}, \bibinfo{person}{Ting Zhang}, {and} \bibinfo{person}{David Lo}.} \bibinfo{year}{2024}\natexlab{}.
\newblock \showarticletitle{Large Language Model for Vulnerability Detection: Emerging Results and Future Directions}. In \bibinfo{booktitle}{\emph{2024 International Conference on Software Engineering (ICSE), New Ideas and Emerging Results (NIER) Track}}. IEEE.
\newblock


\bibitem[Zhou et~al\mbox{.}(2023)]%
        {zhou2022large}
\bibfield{author}{\bibinfo{person}{Yongchao Zhou}, \bibinfo{person}{Andrei~Ioan Muresanu}, \bibinfo{person}{Ziwen Han}, \bibinfo{person}{Keiran Paster}, \bibinfo{person}{Silviu Pitis}, \bibinfo{person}{Harris Chan}, {and} \bibinfo{person}{Jimmy Ba}.} \bibinfo{year}{2023}\natexlab{}.
\newblock \showarticletitle{Large language models are human-level prompt engineers}.
\newblock  (\bibinfo{year}{2023}).
\newblock


\end{thebibliography}










\end{document}